\documentclass[a4paper,aps,twocolumn,showpacs,superscriptaddress,nobibnotes,nofootinbib]{revtex4-2}
\pdfoutput=1

\usepackage{graphicx}
\usepackage{titletoc}
\usepackage{physics}
\usepackage{enumitem}
\usepackage{bbold}
\usepackage{amsmath, amsthm, amssymb,mathrsfs,bm}

\usepackage{mathtools}

\usepackage[normalem]{ulem}
\usepackage{optidef}
\usepackage{algorithmic}
\usepackage{algorithm}

\usepackage{algorithm}
\usepackage{pifont}

\newcommand{\be}{\begin{equation}}
\newcommand{\ee}{\end{equation}}

\usepackage{dsfont}
\usepackage{vwcol}
\usepackage[colorlinks, citecolor =magenta]{hyperref}
\usepackage[usenames,dvipsnames,table]{xcolor}

\usepackage{changepage}

\usepackage[acronym]{glossaries}
\makeglossaries
\newacronym{qkd}{QKD}{quantum key distribution}
\newacronym{di}{DI}{device-independent}
\newacronym{pm}{PM}{prepare-and-measure}
\newacronym{sdp}{SDP}{semidefinite programming}
\newacronym{POVM}{POVM}{Positive Operator Valued 
Measure}
\newacronym{usd}{USD}{Unambiguous state discrimination}
\newacronym{qber}{QBER}{Quantum Bit Error Rate}

\usepackage{float}

\usepackage{tikz}
\usetikzlibrary{
  arrows.meta,
  automata
}
\tikzset{
    -Latex,auto,node distance =1 cm and 1 cm,semithick,
    state/.style ={ellipse, draw, minimum width = 0.7 cm},
    point/.style = {circle, draw, inner sep=0.04cm,fill,node contents={}},
    bidirected/.style={Latex-Latex,dashed},
    el/.style = {inner sep=2pt, align=left, sloped}
}

\usepackage{multirow}
\usepackage{diagbox}

\usepackage{wrapfig}

\usepackage[normalem]{ulem}
\newcommand{\stkout}[1]{\ifmmode\text{\sout{\ensuremath{#1}}}\else\sout{#1}\fi}

\begin{document}

\title{
  Incorrect Citation Association for Articles in Online-Only Springer Nature Journals
}

\author{Tamás Kriváchy}
\email[Contact: ]{tamas.krivachy@gmail.com}
\affiliation{ICFO - Institut de Ciencies Fotoniques, The Barcelona Institute of Science and Technology, 08860 Castelldefels (Barcelona), Spain}

\begin{abstract}
We show that citation metrics of journal articles in many of the online-only Springer Nature journals and associated ones are distorted, going back to articles from 2001. We find that most likely due to an API response error, there are many incorrect references which typically lead to Article Number 1 of a given Volume. Among others, the issue affects journals such as Scientific Reports, Nature Communications, Communications journals, Cell Death \& Disease, Light: Science \& Applications, as well as many BMC, Discovery and npj journals. Beyond the negative effect of introducing incorrect reference information, this distorts the citation statistics of articles in these journals, with a few articles being massively over-cited compared to their peers, while many lose citations; e.g. both in Scientific Reports and in Nature Communications, 5 of the 10 top cited articles have article numbers of 1. We validate the distorted statistics by assessing data from multiple scientific literature databases: Crossref, OpenCitations, Semantic Scholar, and the journals' websites. The issue primarily arises from the inconsistent transition from page-based referencing of articles to article number-based referencing, as well as the improper handling of the change in the publisher's article metadata API. It seems that the most pressing problem has been present since approximately 2011, which we estimate affects the citation count of millions of authors.
\end{abstract}

\maketitle

\section{Introduction}

For better or worse, citation metrics have been a cornerstone of evaluating the impact and relevance of scientific research, particularly since the mid-20th century. The precise number of citations an article receives, i.e. how many other articles reference it, can be difficult to determine exactly and varies from source to source~\cite{doi:10.1086/602160, Li31082010, FRANCESCHINI2016174, FRANCESCHINI2016933, https://doi.org/10.1002/asi.20677, KRATOCHVIL201757, MARTINMARTIN20181160, https://doi.org/10.1002/meet.14504301185, https://doi.org/10.1002/asi.22898, vaneck2019accuracycitationdataweb, Kumpulainen2022, CRL15806}. Regardless of the exact method, having a database without major errors is a crucial element in evaluating the impact of scientific research. The most crucial step in reducing these errors would be to have correct references in research articles, which unfortunately is not always the case~\cite{Key1977, doi:10.2105/AJPH.77.8.1011, doi:10.1177/00220345890680030101, https://doi.org/10.1002/nur.4770100310, 10.1093/ae/43.4.246, deLacey884, https://doi.org/10.1111/j.1547-5069.1998.tb01269.x, Lawson_Fosker_1999, Siebers2000, O'Connor01062001, 00006223-200211000-00006, GOSLING200436, Wyles03052004, Aronsky2005, 00006199-199809000-00010, doi:10.1177/1049731503262131, 10.1108/00012531211244734, Jergas2015}.

Among the many challenges that are encountered when having to handle references, a seemingly simple yet significant one is the transition from paper-based scientific journals to fully online ones. In an attempt to make references clearer and more concise, many online-only journals reference articles via a single ``article number'' instead of using the pages they appear on in the journal. A print-first journal may have a reference such as
\begin{align*}
&\text{Author A, Author B. Fascinating article title.}\\&\text{Nature 777, 8888–8898 (2100)}
\end{align*}
where we used 777 as an example Volume Number, 8888 as the Start Page and 8898 as the End Page and 2100 as the year of publication. In contrast, more modern, online-only journals often use the format
\begin{align*}
&\text{Author A, Author B. Fascinating article title.}\\&\text{Nature Communications 777, 999 (2100)}
\end{align*}
where we used Volume 777 and Article Number 999 as an example.

\begin{figure}[t!]
    \centering
    \includegraphics[width=0.95\linewidth]{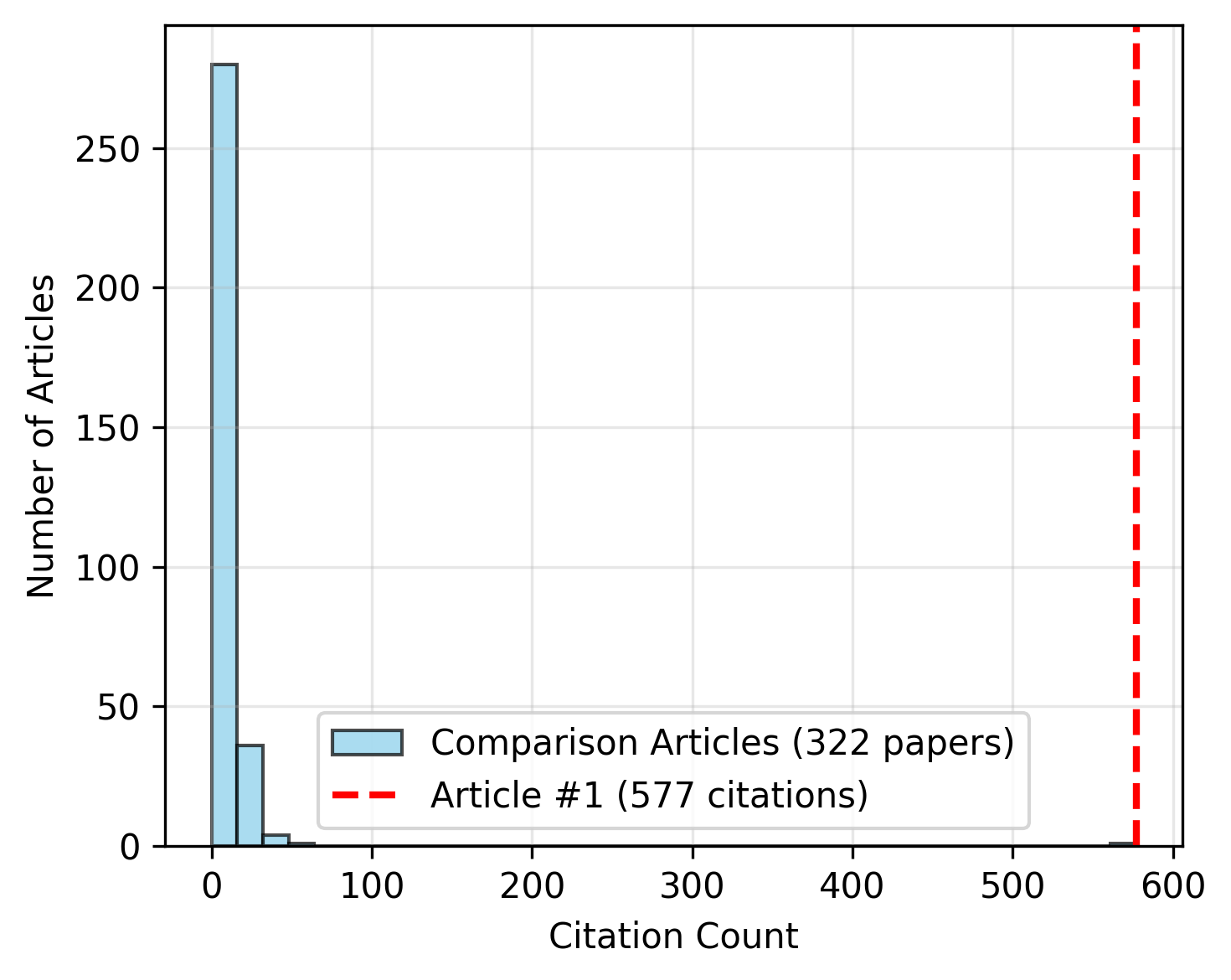}
    \caption{Citation count histogram for Nature Communications Volume 16 (the volume of the current year, 2025) according to Crossref for articles that were published on the same day as Article Number 1 of this volume (data as of 2nd of October 2025). The citation count for Article Number 1 in Vol. 16 is depicted with the dashed red line.}
    \label{fig:fig1}
\end{figure}

Here, we report on a large number of wrongly attributed citations of articles published in journals belonging to or in relation to the Springer Nature Group, which may affect both database handling and correct reference generation. Based on our analysis, the mis-citations happen primarily due to the above adaptation from a page-based numbering to an article number-based one; more specifically, from the improper technical handling of the change. The problem seems to stem from the absence of the Article Number in most formats of the article metadata obtained through the SpringerLink Application Programming Interface (API), or possibly from the handling of the fields in RIS file format provided by the publisher on Springer Nature Link websites. The dominating result is that many times instead of attributing a citation to Article $X$ of Volume $V$, it is attributed to Article $1$ of Volume $V$, though sometimes it may be attributed to another low number, as we detail later. 

Consider, for example, Nature Communications, whose volume numbers have been incremented yearly since 2018. There, the first article to be published in each year typically gets attributed much more citations, whereas the other articles lose those citations; see Fig.~\ref{fig:fig1} for the year 2025 up to here. This effect is robust across various citation count data providers, and is also an issue on the journals' websites themselves. It also seems to be an issue for all online-only journals using article numbers on SpringerLink.

The phenomenon can have countless effects ranging from the inefficient functioning of scientific literature networks, to the improper gauging of scientists' merits, impacting their prestige in the field and how they are judged when it comes to e.g. conference invites, grant and position applications.

The manuscript is structured as follows. In Sec.~\ref{sec:technical} we introduce the technical issues on SpringerLink which we believe could cause problems in citation networks. In Sec.~\ref{sec:citation_effect} we discuss the effects we observed in mistaken citation and distorted citation counts. In Sec.~\ref{sec:scope} we gauge the scope of the problem. Finally, in Sec.~\ref{sec:discussion} we finish with a discussion. In the Appendices, we provide examples that support the findings.

\section{Technical issue}\label{sec:technical}
When exporting a citation from Springer Nature websites, the default and only export option is a file prepared in RIS format. In this format there is no dedicated field for Article Number, hence the Article Number (for those journals which use this) is placed in the Start Page (SP field), each article receives Issue Number (IS) 1 and the End Page (EP) field is not provided in all online-only articles we have seen. Hence, even if one is unaware of the use of Article Numbers, each article may be interpreted as being a single-page digital article starting on page SP of the given Volume. Article numbers are typically incremented by 1, so the subsequently published article will have an Article Number of 1 larger.

The typical user encounters only the previously detailed citation export when using Springer Nature websites. However, if one uses the SpringerLink API provided by the publisher to extract article information in JSON format (one of the most popular formats), one encounters the terrifying fact that the Article Number is not present among the returned values. It is not just a lack of a dedicated field, but the number itself is not present in the API response, as one can verify by reading the response or searching for the specific Article Number. Fields that are available are for example ``volume'', ``number'' (referring to Issue Number), ``startingPage'', ``endingPage''. However, in contrast with the RIS format, ``startingPage'' is not used for the Article Number, but is assigned a value of 1 for each article (in online-only journals),  and ``endingPage'' is assigned the number of pages of the downloaded PDF of the article.

The above described effects clearly create two sources of possible errors within citation management systems.
\begin{itemize}
    \item[\textbf{I.1.}] RIS-based citations contain Article Number as Start Page and provide no End Page. API responses contain Start Page of 1 and End Page which is the length of the PDF.
    \item[\textbf{I.2.}] JSON, JSONP and PAM (XML) format API responses do not contain the Article Number at all, for Meta v1, v2 and Open Access API endpoints.
\end{itemize}
We call these Issues, hence the label I. Naturally, I.2 is much more crucial, as most scientific metadata aggregators use API calls to obtain data. Moreover, the RIS format used in I.1 seems to be used by other publishers as well, but unfortunately leads to an inconsistency between the API and the RIS format (we have not checked other publishers' APIs).\footnote{Note that some publishers prefer to prepend an ``e'' to the Article Number when it is associated with the Start Page field to make a distinction between the two.}

Finally, we note that there \textit{is} a way to obtain the Article Number through the (free tier) official API: the final format offered, JATS formatted XML response \textit{does} contain the Article Number via an ``elocation-id'' tag, as well as in the ``seq'' attribute of the ``issue'' tag (the value of the Issue being 1), and sometimes in ``custom-meta'' tags.

\section{False citation attribution}\label{sec:citation_effect}
\textit{Examples for all observations described here can be found in App.~\ref{app:mistaken_citations}.}

The technical issue described in Sec.~\ref{sec:technical} is one that affects many data sources, including the websites of Springer Nature journals. Here, we summarize two observations (O) in Springer Nature online-only articles which most certainly stem from the issues in Sec.~\ref{sec:technical}.

\begin{figure*}[t!]
    \centering
    \includegraphics[width=0.32\linewidth]{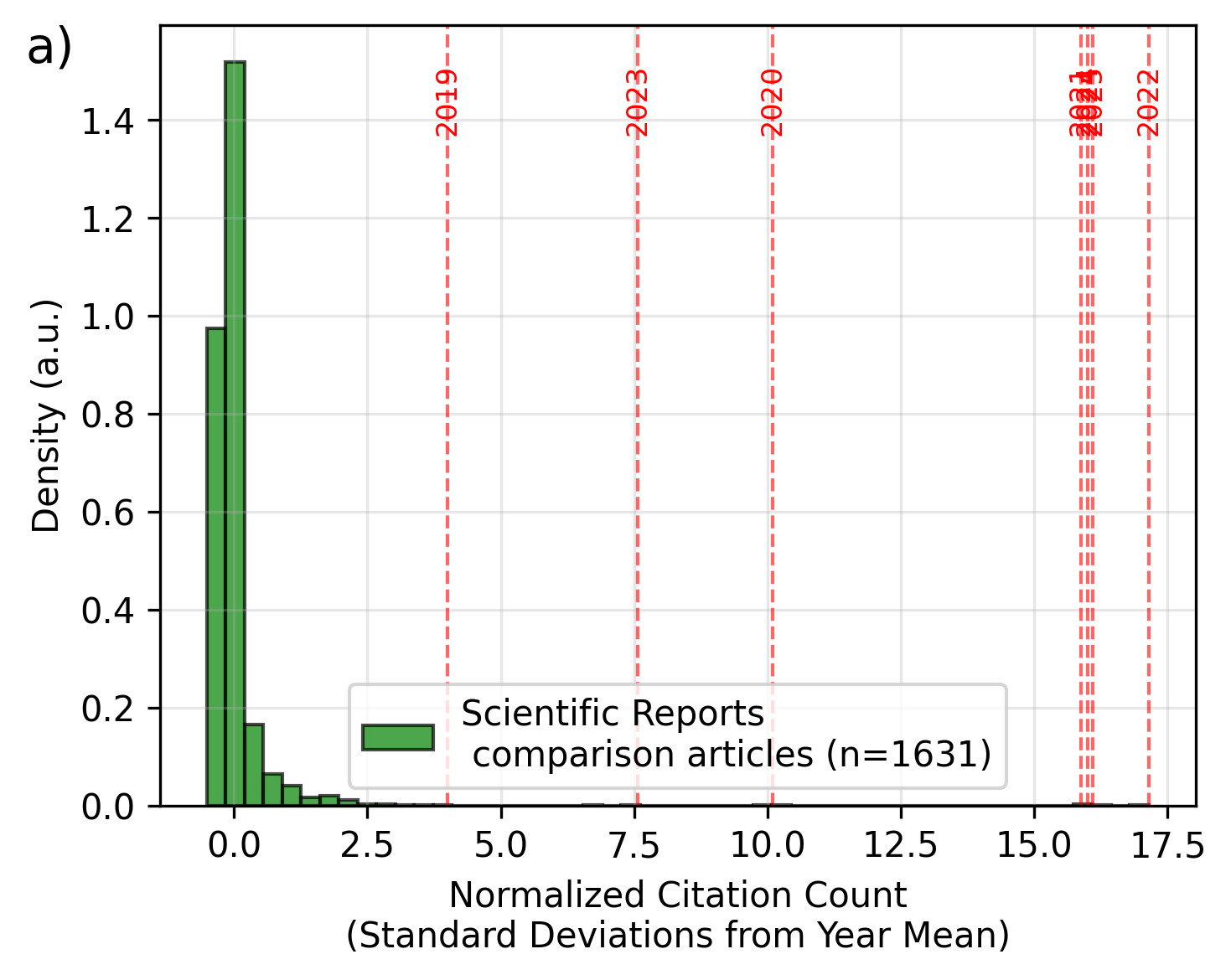}
    \includegraphics[width=0.32\linewidth]{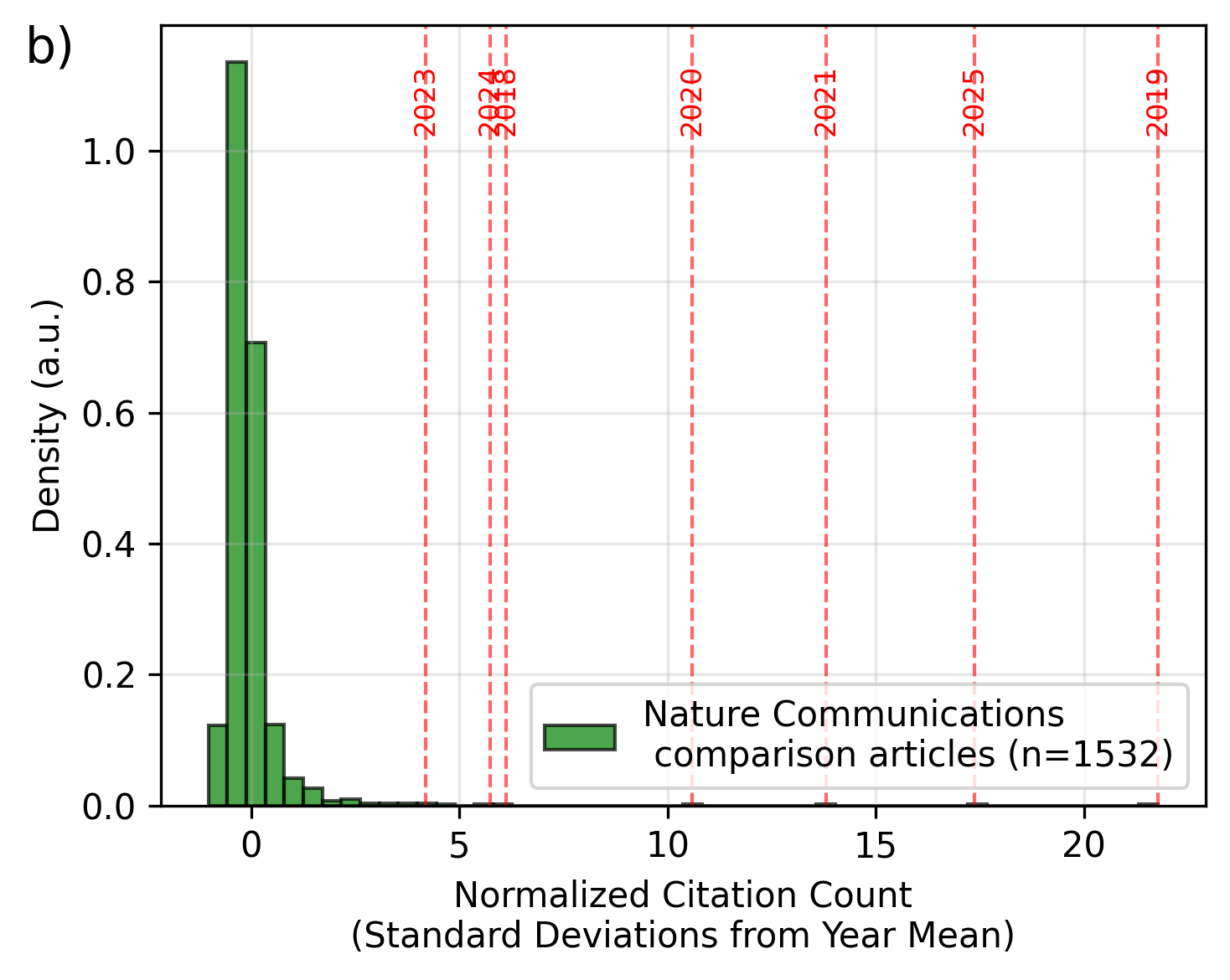}
    \includegraphics[width=0.32\linewidth]{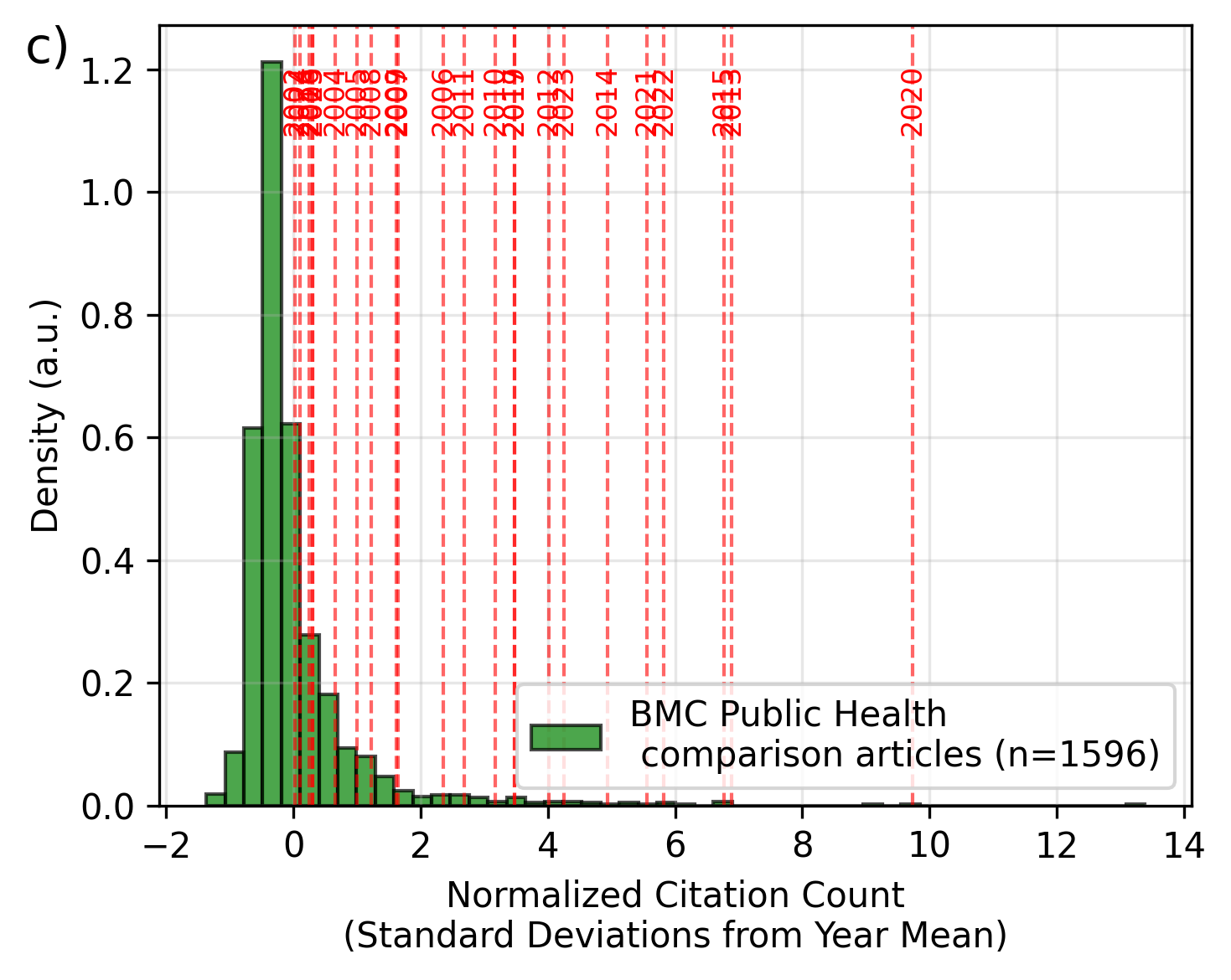}
    \caption{Citation count histogram for a) Scientific Reports b) Nature Communications and c) BMC Public Health according to Crossref for articles published on or near the day of Article Number 1's publishing date for the years a) 2018 to 2025 b) 2019 to 2025 c) 2002 to 2025 (2022 excluded for Nature Communications, since Article 1 was a correction; citation count as of 2nd of October 2025). The citation counts for Article Number 1s for each year are depicted with dashed red lines. Citation counts are normalized among years such that the mean is 0 and standard deviation is 1, so they can be plotted together. For other citation count sources and unnormalized data plots, see App.~\ref{app:histograms}.}
    \label{fig:fig2}
\end{figure*}

\begin{itemize}
    \item[\textbf{O.1.}] False attribution of a citation to Article $X$ of Volume $V$ to Article $1$ of Volume $V$. This most likely happens when the SpringerLink API is used, and the ``startingPage'' (value of 1 for all articles) is used instead of (the not provided) Article Number. Due to the reliance on the API, this is an issue for several scientific article metadata providers.
    \item[\textbf{O.2.}] False attribution of a citation of Article $X$ of Volume $V$ with a PDF of length $L$ to Article $L$ of Volume $W$ ($W\neq V$ or $W=V$ are both possible) of the same journal. This was observed on SpringerLink websites, where the reference's title was correct, however the journal reference was incorrectly
    \begin{equation*}
        \text{Journal Name }V(1):L
    \end{equation*}
    (most likely the (1) stems from the Issue Number), or
    \begin{equation*}
        \text{Journal Name }V:L
    \end{equation*}
    Note that the journal itself recommends articles to be cited as 
    \begin{equation*}
        \text{Journal Name }V, \,X,
    \end{equation*}
    hence it does not follow its own recommended citation style. 
    Furthermore, the hyperlink which should lead to the referenced Article leads to the incorrect reference Article $L$ of Volume $V$. For other references we found that the reference was written incorrectly, but the hyperlink pointed to the correct article. Note that in the examined cases the Google Scholar link provided alongside the Article hyperlink points to the correct article. 
    \item[\textbf{O.3.}] Incorrect article title, journal name, hyperlinks for certain articles in the References section. The origin of the problem is unclear.
\end{itemize}
At this point the reader may wish to go to the website of one of these journals to search for an Article 1 in order to see whether it really is cited by articles from different fields by mistake. Unfortunately this is not as easy as it sounds, as on the Nature Portfolio advanced search, there is a field ``volume'' and another one ``start page / article no''. By selecting a journal, writing in some volume and a 1 in the latter field to get article number 1, we actually retrieve all articles for the given volume, as they all have start page 1. Note that for any other article number the search seems to work, it is only article 1 that cannot be found easily. A work-around on the website is to manually check all articles from the first day of publication of a Volume. For certain journals this means having to click through hundreds of articles. In the Supplementary Data~\cite{data_of_this_article} we provide several Article 1 references which we extracted via the API.

\begin{figure*}[t!]
    \centering
    \includegraphics[width=0.99\linewidth]{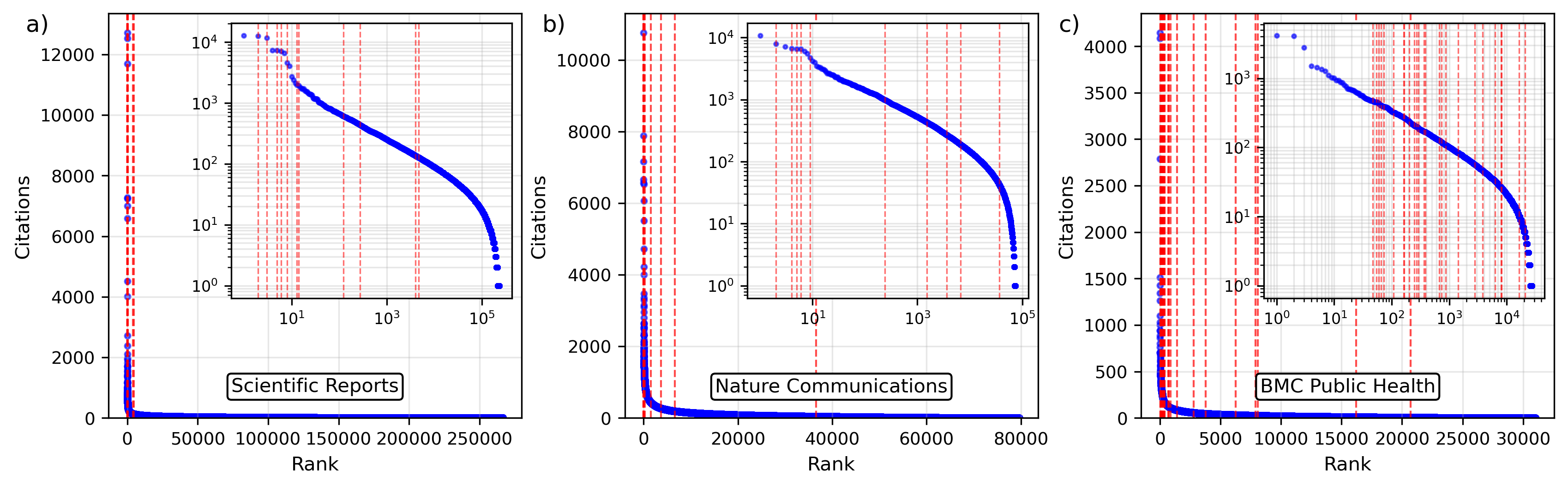}
    \caption{Citation counts ranked for all published articles in a) Scientific Reports b) Nature Communications and c) BMC Public Health according to Crossref (data from approximately 2nd of November), with a log-log plot of the same data in the insets. Article 1 ranks are marked with dashed red lines. In both Scientific Reports and Nature Communications, 5 of the top 10 cited articles have article numbers of 1.}
    \label{fig:fig3}
\end{figure*}

Whereas we only verified O.2 and O.3 through manual checks, we could easily verify the impact of O.1 in two ways. First, by comparing the citation count of articles with Article Number 1 to articles with a similar age (comparison articles). Comparison articles of a given Article 1 are those that were published on the same day in the same journal. If we found less than 15 comparison articles, then we included subsequent days until we had at least 15 comparison articles. For example in Fig.~\ref{fig:fig1} we found the publication date of Article 1 of Nature Communications Vol. 16 (the Volume of 2025), which was 2nd of January 2025. Using the Crossref API we extracted the citation count for 322 articles published on the same day in the same journal. We checked three different journals (Scientific Reports, Nature Communications, BMC Public Health, three of the largest journals based on article volume in recent years) and four citation count providers (Crossref, OpenCitations, Semantic Scholar and the citation numbers given on the journal's own website). Detailed histograms can be found in App.~\ref{app:histograms}. Here, besides the explicit example in Fig.~\ref{fig:fig1}, we provide an aggregate plot from multiple years for the three journals using the Crossref citation count in Fig.~\ref{fig:fig2}. In order to adjust for variability among the age of publication and journal type, we normalize the citation count for each journal and year such that they have a mean of 0 and a standard deviation of 1. As such, they can be visualized and compared on a single plot.

The second verification method consists of examining the position of Article 1s in the citation count rankings for the three journals studied in detail, Scientific Reports, Nature Communications and BMC Public Health. As the four citation count sources resulted in similar results for the histogram studies, we used only Crossref to extract the citation count of all articles in the three journals. We can see citation count vs. ranking in Fig.~\ref{fig:fig3}, with Article 1 ranks (from different years) marked with vertical dashed red lines. If O.1 would not exists, we would expect a more or less uniform distribution of Article 1 ranks on a linear scale. On the contrary, we observe an extreme concentration of Article 1 ranks. For both Scientific Reports and Nature Communications, 5 of the 10 most cited articles are Article 1s. For BMC Public Health, the concentration is less significant, but still we find 5 Article 1s in the top 100 articles, 10 in the top 300, and all but two Article 1s in the top 10,000 articles (the exceptions being the most recent Article 1s, from 2024 and 2025), in a journal that has 31,000 lifetime articles.

The histograms (Fig.~\ref{fig:fig2} and App.~\ref{app:histograms}) as well as the rank plots (Fig.~\ref{fig:fig3}) reveal the dramatic distortion of citation counts. In the histogram plots, for almost all years, journals and citation count sources, Article Number 1s have an incredible number of citations, forming outliers in the dataset. In the rank plots, we see Article 1s overrepresented among the most cited papers. Note that we expect a similar, though perhaps less drastic effect with article numbers that are the same as typical research article page lengths, such as $4, 5, \dots \sim 20$, due to O.2. We have not gathered statistics on this effect, as we decided to focus on Article Number 1s. Note that there is some variation among journals, with the effect being more moderate in BMC Public Health. Though even for this journal not a single Article 1 citation count went below the mean citation count in Fig.~\ref{fig:fig2}

\section{Scope of the problem}\label{sec:scope}
Springer Nature Group is among the largest scientific publishers both in terms of number of journals (over 3000) and number of articles published per year (over 482,000 in 2024), thus it is worth considering the wider impact of this problem. 
We examine its extent in the following aspects: affected journals, years, articles, authors, data sources, and journal metrics. Finally, we conclude with a comment on other publishers.







\textbf{Journals} Springer publishes and shares content on the following websites: link.springer.com, nature.com, BioMedCentral.com, SpringerOpen.com~\cite{nature_info_website_names}. On all of them we found examples for I.1 and I.2.
We have verified manually that many journals are affected by I.1 and I.2, such as Scientific Reports, Nature Communications, BMC journals, Communications journals, Discovery journals, npj series\footnote{Note that some of these journals changed their article numbering protocol throughout the years, affecting the detailed effects of these issues. For example, Nature Communications began publishing in 2010. Up to year 2017 the Volume Number was incremented each year, however Article Numbers were not reset to 1 at the start of a new Volume (year). Hence from 2010 to 2017 there was only one Article Number 1. Starting from 2018, Article Number was reset to 1 at the beginning of each year, hence from 2018 each year exhibits O.1.}. Due to API limitations and a lack of a complete database of journals, we could not provide a full list of journal identifiers and affected years. A list off all journals for which we tested I.2 can be found in App.~\ref{app:journals}. Recall, journals are affected if they have article number-based referencing of articles - which it seems is the case for most newer journals, those that are online-only. Journals that have been running for a long time or which have a low number of articles published, such as Nature and the Nature series, typically use page numbering or no numbering at all, meaning that they are presumably unaffected by I.1 and I.2.

\begin{figure*}[t!]
    \centering
    \includegraphics[width=0.48\linewidth]{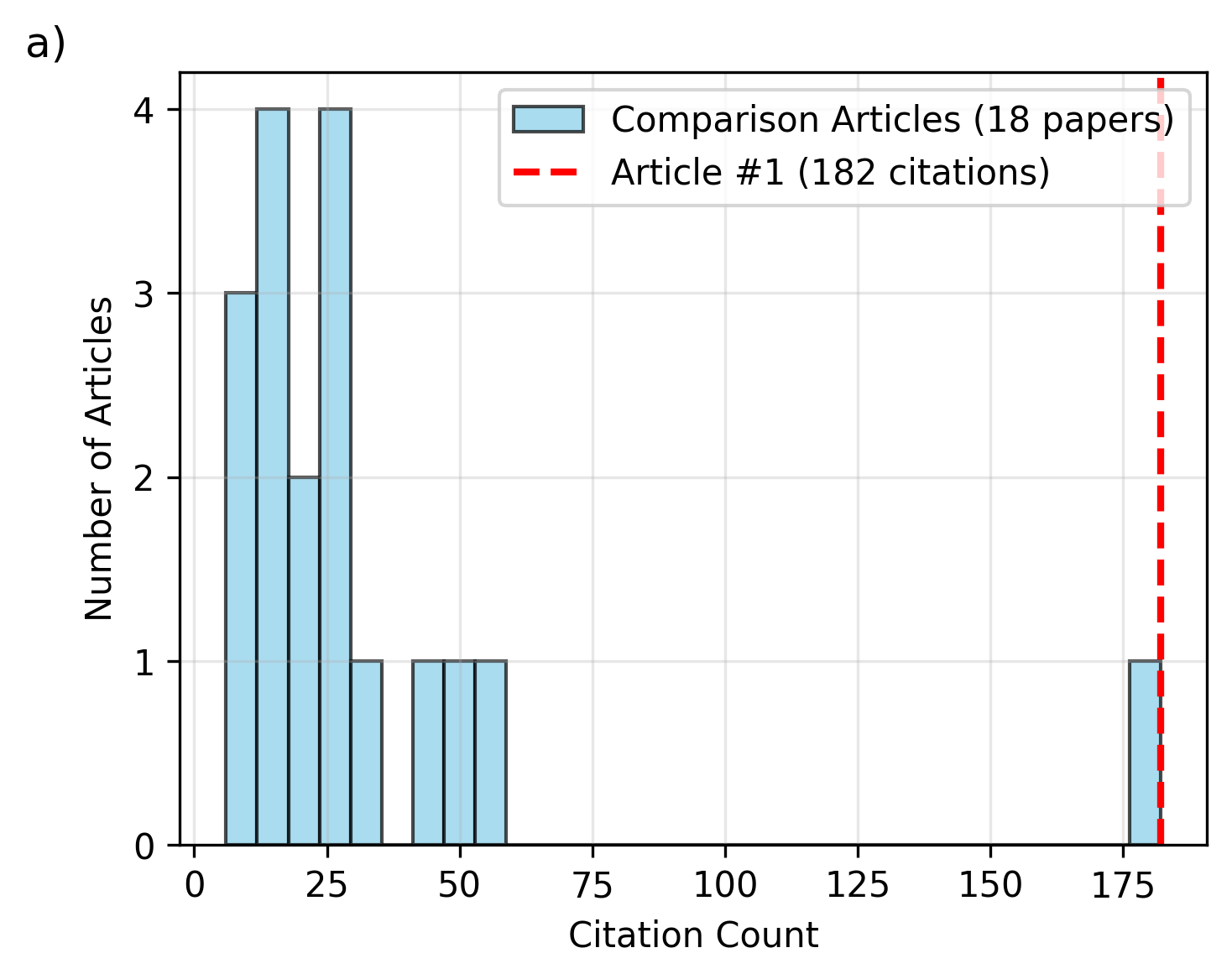}
    \includegraphics[width=0.48\linewidth]{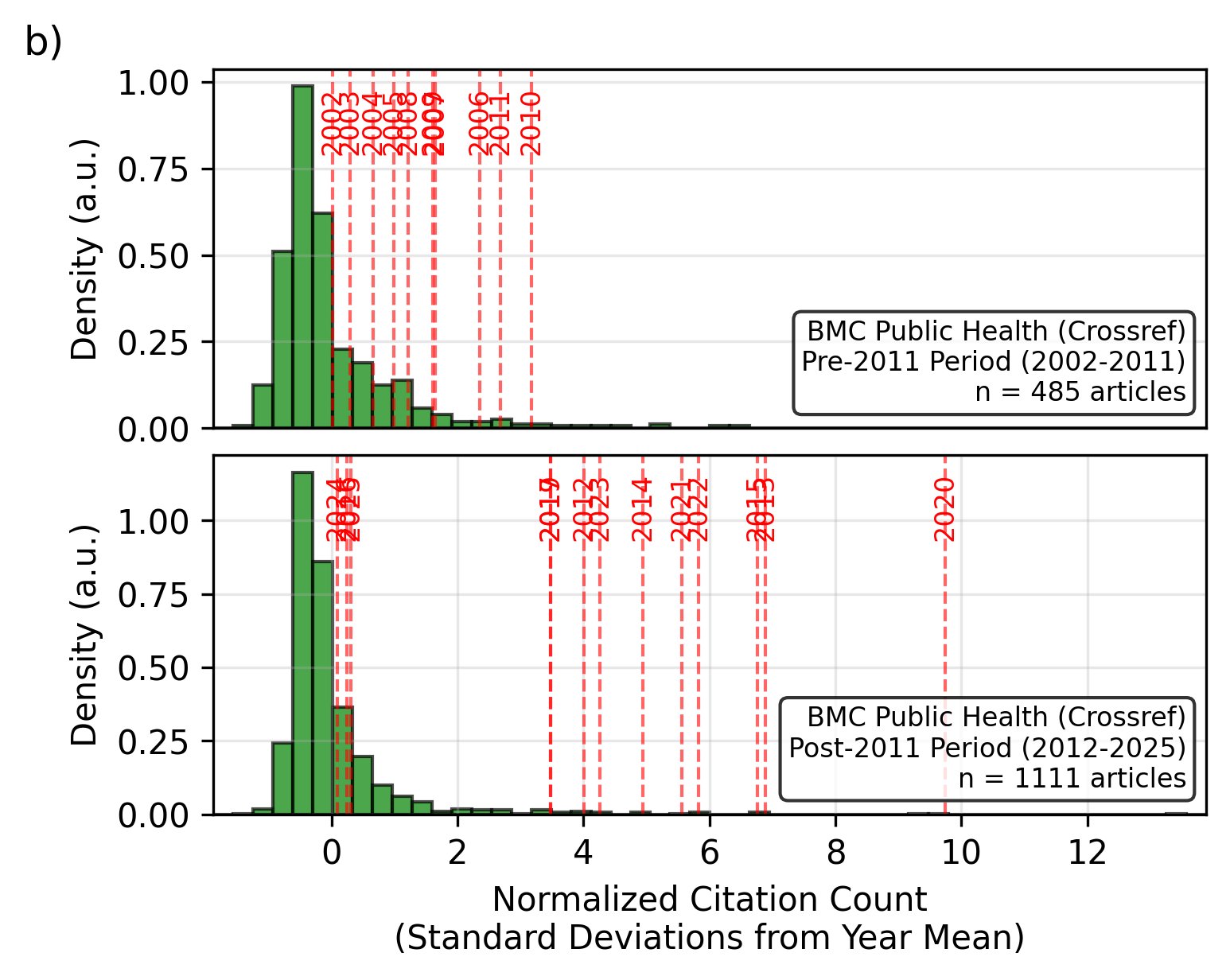}
    \caption{Evidence of temporal extent of issues. a) Already in 2020, the citation count histogram of Nature Communications Vol. 3 was distorted. Citation count taken from archived webpages of the journal's website using the Internet Archive's Wayback Machine. Citation count of Article 1 was taken from 2020 September 22, and of comparison articles the from the first available 2021 webpage archive. b) Normalized citation count histogram for BMC Public Health (data from Crossref) plotted in two parts: pre-2011 and post-2011, the year when the SpringerLink API was introduced. For the post-2011 years, the normalized citation count for Article 1s was only as low as in the pre-API era for years 2016, 2024 and 2025. Similar effect observed for other data sources (see App.~\ref{app:histograms}).}
    \label{fig:fig4}
\end{figure*}

\textbf{Years} In the years affected we have to make a distinction between the years of published articles which have been affected, which range back to the year 2000 (e.g. some BMC journals) versus the years that the SpringerLink API has operated without returning an Article Number (I.2). It is difficult to obtain information about the latter, as API calls cannot be made into the past. Some relevant dates, one of which could be the point that I.2 began: SpringerLink has been running since 1996, launched its public API in 2011~\cite{history_1_API1,history_2_API1}, refined it in 2012~\cite{history_3_API2}, and modified it in 2015 at the time of the Springer Nature merger. We find three significant pieces of evidence of the temporal extent of I.2 with the help of the Internet Archive Wayback Machine and using the citation data at hand.
\begin{itemize}
    \item (At least 3 years old) We found archived versions of an Article 1 which has the ``This article is cited by'' section with article references which \textit{do not} actually cite that article. The ``This article is cited by'' feature on the website seems to have been introduced in 2022 or before (at least for Nature Communications, where we checked), which is the year in which we observed O.1~\cite{wayback_2022_10_16_article1_2018}. Hence, O.1 (and O.2, I.2 probably) has been around for at least 3 years.
    \item (At least 5 years old) Again, using the Internet Archive, we gathered citation counts from the journal's archived websites from the year 2020. Specifically, we took the Article 1 and Comparison Articles from Nature Communications Vol. 3 (year 2012), and extracted their citation count as it appeared on the journal website in approximately 2020 (data from 2020 September 22 for Article 1, and earliest 2021 available data for all comparison articles), and plotted the histogram in Fig.~\ref{fig:fig4}. The figure clearly exhibits the same trend as we can observe today.
    \item (Most probably 14 years old) Given the prevalence of the principle of backward compatibility in API systems, we expect that fields and tags in API responses have not been removed throughout the years, hence it is easily possible that the issue exists since the early days of the public API, i.e. for a duration of $\sim$14 years. Indeed, looking at the yearly histograms for BMC Public Health in App.~\ref{app:histograms}, we see that O.1 seems to be much more prevalent in the years after 2011, which happens to be the year that the API was introduced. To illustrate this, we plot two aggregate citation count histograms in Fig.~\ref{fig:fig4}, one for volumes up to 2011, and one for volumes after. We observe much larger typical standard deviations for the ones after 2011. This may be due to more correct references before 2011 or to cached citation count data from the pre-API era.
\end{itemize}
Finally, note that we found references to I.1 on the Zotero forum from the year 2020~\cite{zotero_forum_1,zotero_forum_2}.

\textbf{Articles} As we had difficulties collecting the exact list of journals which are affected, we cannot provide an exact count. Note, however, that the two largest journals based on number of articles per year, Scientific Reports and Nature Communications, are affected, as well as the BMC journals, which comprise a large number of high-volume journals using article number referencing. The total number of articles for Scientific Reports is  $\sim$250,000, for Nature Communications $\sim$75,000, and for several BMC journals and Discover Applied Sciences\footnote{BMC Public Health $\sim$ 30,000, BMC Cancer $\sim$ 19,000, BMC Genomics $\sim$ 18,000, BMC Health Services Research $\sim$ 17,000, BMC Infectious Diseases $\sim$ 15,000, Trials $\sim$ 11,000, BMC Pregnancy and Childbirth $\sim$ 9,000, Discover Applied Sciences $\sim$ 7,000.} $\sim$126,000. So only for these 10 journals there are about 450,000 potentially affected articles, with the total number probably even higher. Springer Nature claims to host 7 million articles~\cite{nature_info_website_names}. Given the immense growth in online articles in the past years, one can expect that a significant proportion of the 7 million are in online-only journals, putting the real number of affected articles to be in the millions.

\textbf{Authors} In the three journals studied in detail, we counted approximately 1.5 million unique authors, with a mean author count of 4-6, depending on the journal. Given that besides the articles in these three journals there must be at least multiple hundreds of thousands of affected articles, we expect there to truly be multiple millions of authors affected.


\textbf{Journal metrics} (\textbf{O.1}) Certain metrics, such as Source Normalized Impact per Paper (SNIP), could be sensitive to mistaken citations within the same volume (O.1), particularly in the case of multidisciplinary journals such as Scientific Reports and Nature Communications. SNIP weighs citation count based on the field's total citation count. Since multidisciplinary journals publish articles from multiple fields, the specific field of the article that happens to be Article Number 1 will significantly influence the calculation. For example, if an article from a field which typically has low citation count becomes Article Number 1, then it will receive mistaken citations proportional to the number of citation counts from high-citation-count-fields in the same journal, leading to distorted source normalized metrics. (\textbf{O.2}) Though the most common effect seems to be mistaken citations within the same journal and within the same year, we have observed mistaken citations where different volume numbers appear (O.2). It is unclear how often this occurs, but due to the cross-over between different volumes (which often correspond to years) it does impact temporally sensitive aggregate metrics of journals such as Impact Factor (IF). (\textbf{O.3}) We have not studied the extent of this problem, but there seem to be references to articles in the incorrect journals (O.3), leading to a shift of citation counts from one journal to another. (\textbf{I.1, I.2}) Furthermore it could be that sometimes due to missing Article Number or improper reference data that was generated due to I.1 or I.2, certain citations are simply missed and not counted, further modifying such metrics. 

\textbf{Data sources} Finally, we deduce from the observed outlier distribution that many data sources have this issue. In particular we examined Crossref, OpenCitations, Semantic Scholar and the citation numbers given on the journals' own websites. They all had Article Number 1s as citation count outliers. Moreover, the journals' own websites contain incorrect articles in their ``This article is cited by'' sections. Most likely all of these use the SpringerLink API and hence encounter I.2. Furthermore, we found examples of Google Scholar also missing a citation perhaps due to I.2, as well as including incorrect articles in ``cited by'' lists. Unfortunately it was more difficult to check statistics from sources such as Web of Science and Scopus, as they have limited free access. Both allow for author searches, where we find great variation between citation count both for authors who appeared in Article 1s and those who appeared on comparison articles. Thus, the extent of the issue for these sources is unknown.

\textbf{Publishers} We have checked other publishers Elsevier, Wiley and MDPI to see whether they have I.2 as an issue. Elsevier has no free API, so we could not check it directly. Wiley doesn't have an API specifically for metadata, only full-text PDF downloading, and MDPI refers those interested in APIs to use Crossref. Using Crossref, we could access article information for sample articles from all of the above publishers (Elsevier, Wiley, MDPI, Springer Nature), and found that Crossref returns a dedicated Article Number field for all of them, \textit{including} for articles from Springer Nature. 


\section{Discussion}\label{sec:discussion}

The issues uncovered here have multiple implications. On the scientific side, the mistaken citations reduce the efficiency of finding relevant literature, for example finding articles that build on a given researcher's work. On the scientific impact evaluation side, the issues distort metrics based on an article's or researcher's citation network. Perhaps the most crucial aspect is the number of citations an article/researcher receives, an aggregate metric that is commonly used.

Misjudging scientific impact based on inaccurate citation counts can have quite deep implications. At first glance, for the millions of scientists whose citation counts were reduced, the impact may seem minor, as it amounts to just a few citations per researcher. However, we believe the larger impact may be in the overestimation of the importance of Article Number 1s, and the merits of their authors. The competition for funding and permanent positions can be fierce, and though the decisions are not as simple as assessing citation counts of applicants, it may be that the impressively large citation counts of certain applicants biased decisions in an unfair way. Note that even if the bug is fixed and Article 1 citation metrics are more on-par with their peers' ones, we expect that they will forever be cited more on average. This is due to the simple effect that articles with a high number of citations seem as more legitimate references, and are thus typically cited more. The extra legitimate citations will remain with us even after the problem is fixed.

What next? Clearly, for the benefit of the scientific community, Springer Nature must fix I.2 as soon as possible, notify the most frequent users of the API, and analyze the extent of the issue as well as possible, using internal resources inaccessible to us. In particular, the most important question would be: since when has the Article Number been missing from API responses? Since when are citation counts distorted? Which journals are affected by improper citations due to O.1 and O.2? Why have mistakes such as O.3 happened and how many journals/years/articles are affected by it? We believe the scientific community would deserve a public report on this, as well as transparent strategy from Springer Nature on how it will fix and mitigate the issue as well as possible. Moreover, a push from Springer Nature to have a more unified citation metadata format and numbering may be beneficial, for example fixing the inconsistency coming from I.1. Furthermore, scientific citation network aggregators such as Crossref, Semantic Scholar, OpenCitations, etc., must update their databases promptly after the issue is fixed. Even with an instantaneous fix of I.2, it could take quite some time for citation networks to be reevaluated, databases to be updated. These companies should issue information about the completion of the updates. At this point journal metrics should also be reevaluated.

Finally, we should all be attentive, so future issues like this do not persist for years and years.

\emph{Note added -- } The manuscript has been sent to Springer Nature, who are in the process of assessing the observations.

\section{Epilogue}
During my doctoral studies, I was somewhat surprised when one of my supervising professors, Nicolas Gisin, returned a draft of a publication with some comments. There were not many comments, but one was to review the references! He had found a few errors in names (accents), some preprints which have since been published, etc. A well-established, experienced scientist who has many students and much more important things to do, was going through the reference list of the draft of my manuscript. Why? It was surprising to me, someone who grew up in a more digital, automated age, where we are used to trusting digital tools to take care of tasks instead of us. Why does anyone need to look through reference lists? I used reference management software to generate them; clearly, everything should be cited properly... 

Of course, my professor was right to do so; by today, I wish to believe that I have learned my lesson, as the current issues were also uncovered due to manual verification of machine-generated reference information. Perhaps the moral here is to periodically check on the machines, just to be sure everything is functioning as we intend.

\section{Data availability}
We provide data at Ref.~\cite{data_of_this_article} containing raw API responses from SpringerLink at the time of writing, as well as code and datasets needed to reproduce the findings.

\section{Acknowledgments}
This research could not have been done without the publicly available resources of Nature Springer, Crossref, Semantic Scholar and OpenCitations. We believe that transparency in science is important and these tools are important enablers for this. We are grateful for strategic input from Antonio Acín, Nicolas Gisin, Marcus Huber and Nicolas Brunner. We acknowledge further insightful discussions with Donato Farina, Gabriel Senno and Mohammad Mehboudi. Finally, we are grateful to the scientists who reproduced the currently observed issues on their own devices: Paul Erker, Sébastien Designolle, Gabriel Senno, Alejandro Pozas-Kerstjens, Giacomo Franceschetto, Emanuel-Cristian Boghiu, Dániel Barabási and Flavien Hirsch. We acknowledge funding from the Swiss National Science Foundation (project 214458).

\newpage
\onecolumngrid
\newpage
\appendix

\section{Examples for mistaken citation}\label{app:mistaken_citations}

\subsection{Example of O.1 from the publisher's website in 2022}
We give an example of O.1, i.e. the observation that an Article 1 of volume $V$ is referenced instead of an Article $X$ of volume $V$. We use the journal's website to show this, specifically from the year 2022. We then compare it to today's version, which also exhibits this error.

\begin{figure}[b!]
    \centering
    \includegraphics[width=0.85\linewidth]{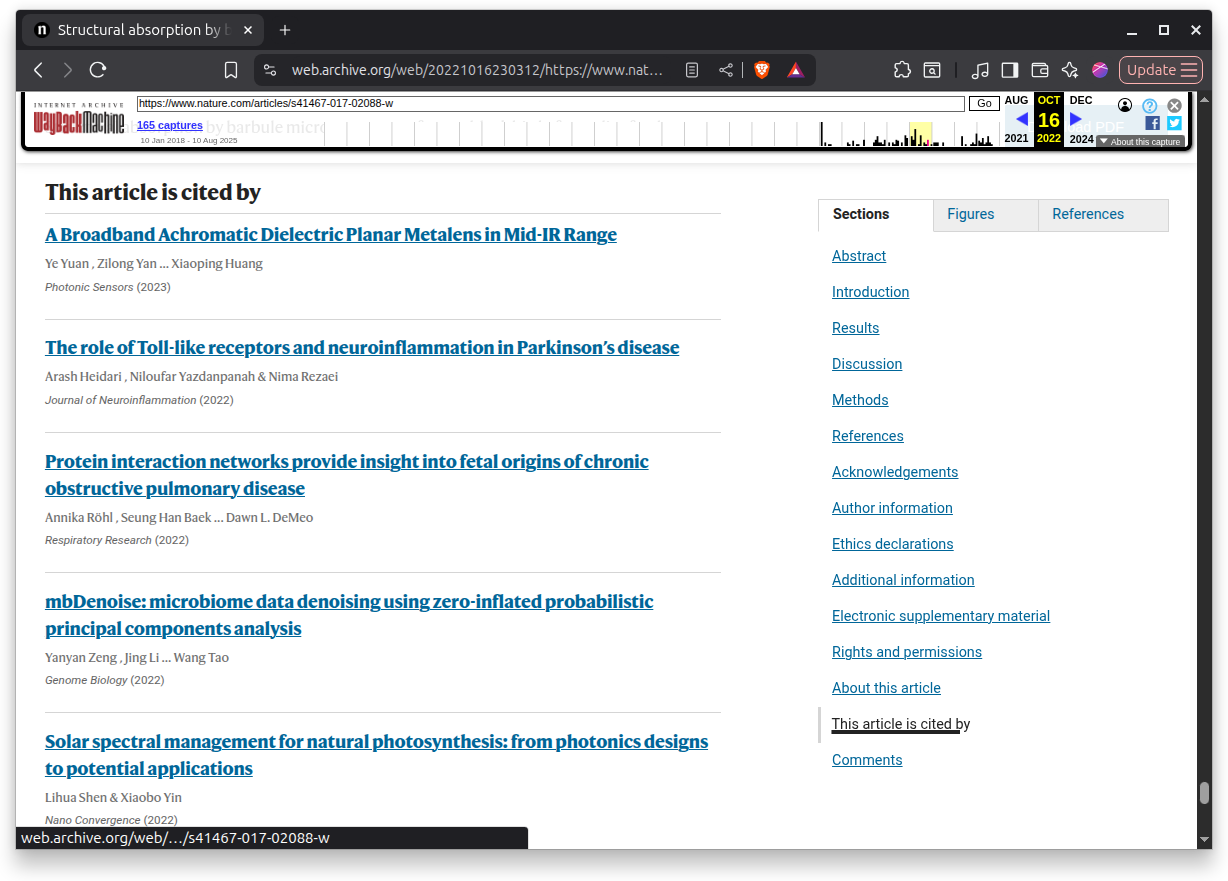}
    \caption{Screenshot of the webpage of Article 1 from 2018 in Nature Communications (titled \textit{Structural absorption by barbule microstructures of super black bird of paradise feathers}), archived on 16th of October 2022. None of the articles in the ``This article is cited by'' section actually cite the current article. This can immediately be suspected from the titles, but it can also be verified manually.}
    \label{fig:app:wayback2022}
\end{figure}

Using the Internet Archive Wayback Machine we found a version of the article website of
\begin{align*}
    &\text{McCoy, D.E., Feo, T., Harvey, T.A. et al. Structural absorption by barbule microstructures}\\
    &\text{of super black bird of paradise feathers. Nat Commun 9, 1 (2018).}
\end{align*}
archived on the 16th of October 2022. By going to the ``This article is cited by'' section as in Fig.~\ref{fig:app:wayback2022}, we find five article references, of which none actually cite the above article. They do, however, cite articles from the same Volume 9. For easy verification by the reader we provide the DOIs of these articles here.
\begin{itemize}
  \item \href{https://doi.org/10.1007/s13320-022-0667-4}{10.1007/s13320-022-0667-4}
  \item \href{https://doi.org/10.1186/s12974-022-02496-w}{10.1186/s12974-022-02496-w}
  \item \href{https://doi.org/10.1186/s12931-022-01963-5}{10.1186/s12931-022-01963-5}
  \item \href{https://doi.org/10.1186/s13059-022-02657-3}{10.1186/s13059-022-02657-3}
  \item \href{https://doi.org/10.1186/s40580-022-00327-5}{10.1186/s40580-022-00327-5}
\end{itemize}

\begin{figure}[t!]
    \centering
    \includegraphics[width=0.85\linewidth]{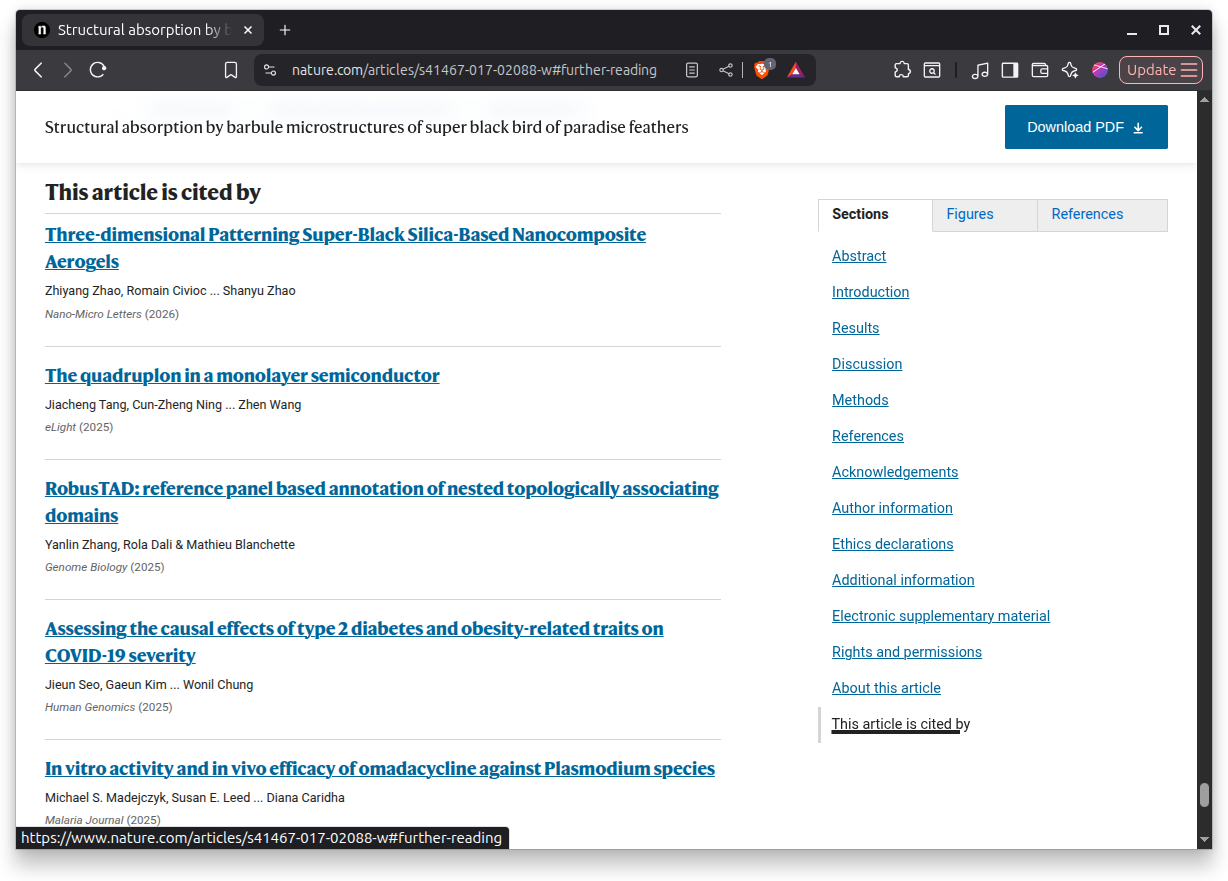}
    \caption{Screenshot of the webpage of Article 1 from 2018 in Nature Communications (titled \textit{Structural absorption by barbule microstructures of super black bird of paradise feathers}), taken 2nd of October 2025. One of the articles in the ``This article is cited by'' section actually cites the current article, the rest do not.}
    \label{fig:app:wayback2025}
\end{figure}

We also took a screenshot of the article on the 2nd of October 2025, visible in Fig.~\ref{fig:app:wayback2025}. Among the 5 articles provided in the ``This article is cited by'' section, only four are incorrect. Again, we provide the DOIs here.
\begin{itemize}
  \item \href{https://doi.org/10.1007/s40820-025-01870-6}{10.1007/s40820-025-01870-6} (Correct)
  \item \href{https://doi.org/10.1186/s43593-025-00081-1}{10.1186/s43593-025-00081-1}
  \item \href{https://doi.org/10.1186/s13059-025-03568-9}{10.1186/s13059-025-03568-9}
  \item \href{https://doi.org/10.1186/s40246-025-00747-4}{10.1186/s40246-025-00747-4}
  \item \href{https://doi.org/10.1186/s12936-025-05448-w}{10.1186/s12936-025-05448-w}
\end{itemize}

\begin{figure}[t!]
    \centering
    \includegraphics[width=0.75\linewidth]{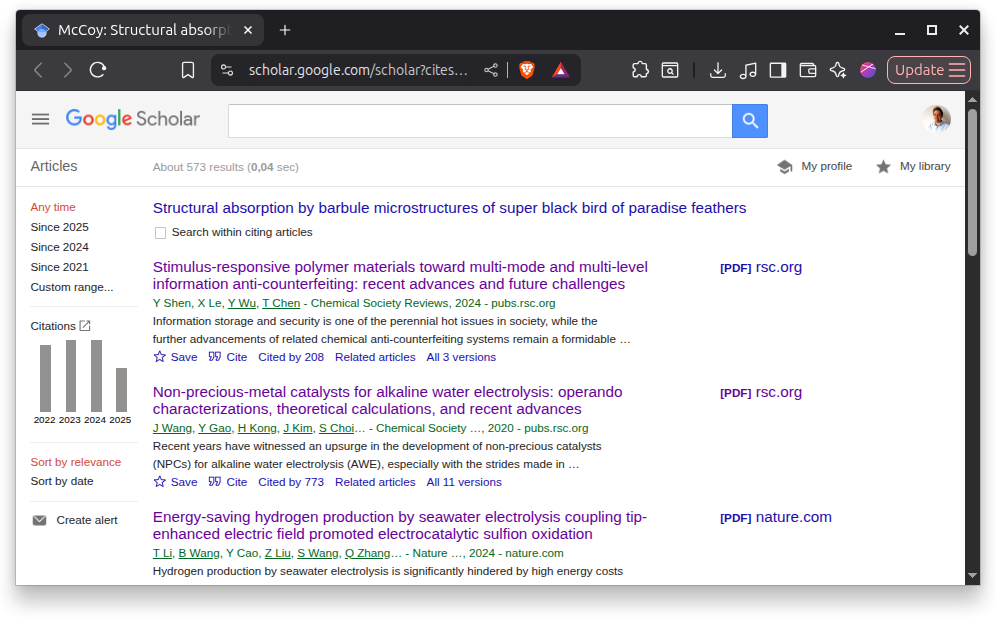}
    \caption{Screenshot of the Google Scholar ``Cited by'' page for Article 1 from 2018 in Nature Communications (titled \textit{Structural absorption by barbule microstructures of super black bird of paradise feathers}). Already the second recommended article does not actually cite the article under examination.}
    \label{fig:app:googlescholarBAD}
\end{figure}

There are major differences in which citation count source one uses to gauge this Article's citation count. For example, at the time of writing the webpage claims over 7400, Crossref 6476, OpenCitations 7181, Semantic Scholar 5279, and Google Scholar 573 citations. The more moderate amount of citations of Google Scholar may be due to a more text-centered (search-based) approach of linking articles. Unfortunately, this source is also not completely reliable for the true citation count. For example, when checking the articles that cite this article (Fig.~\ref{fig:app:googlescholarBAD}, already the second recommended article \textit{does not} actually reference it. It is
\begin{quote}
    Jian Wang, Yang Gao, Hui Kong et al. Non-precious-metal catalysts for alkaline water electrolysis: operando characterizations, theoretical calculations, and recent advances, Chem. Soc. Rev., 2020, 49, 9154 \href{https://doi.org/10.1039/D0CS00575D}{https://doi.org/10.1039/D0CS00575D}
\end{quote}

\newpage
\subsection{Example of O.2}\label{app:exampleO2}

Consider the article
\begin{quote}
Xu, W., Yi, S., Liu, J. et al. Nitrile-aminothiol bioorthogonal near-infrared fluorogenic probes for ultrasensitive in vivo imaging. Nat Commun 16, 8 (2025). \href{https://doi.org/10.1038/s41467-024-55452-y}{https://doi.org/10.1038/s41467-024-55452-y}
\end{quote}
Note that it is \textit{not} an Article 1, but an Article 8. In the ``This article is cited by'' section, at the current time we find one article~\cite{wayback_2025_10_03_O2example_2025},
\begin{quote}
    Jia, Z., Qi, P., Wang, J. et al. Molybdenum disulfide supported on chitin carbon aerogels as an efficient and stable hydrogen evolution electrocatalyst. Ionics 31, 2701–2714 (2025). \href{https://doi.org/10.1007/s11581-025-06089-4}{https://doi.org/10.1007/s11581-025-06089-4}
\end{quote}
This article references two Nature Communications articles in the following format
\begin{quote}
    18. Xu J, Shao GL, Tang X, Lv F, Xiang HY, Jing CF, Liu S, Dai S, Li YG, Luo J, Zhou Z (2022) Frenkel-defected monolayer MoS2 catalysts for efficient hydrogen evolution. Nat Commun 13(1):8
\end{quote}
\begin{quote}
    19. Zheng ZL, Yu L, Gao M, Chen XY, Zhou W, Ma C, Wu LH, Zhu JF, Meng XY, Hu JT, Tu YC, Wu SS, Mao J, Tian ZQ, Deng DH (2020) Boosting hydrogen evolution on MoS2 via co-confining selenium in surface and cobalt in inner layer. Nat Commun 11(1):10
\end{quote}
The correct article number would be 2193 and 3315, respectively. The incorrect article numbers are actually the length of the PDFs of the articles: 8 pages and 10 pages. The (1) refers to the Issue Number. Hyperlinks are conveniently provided with these articles to the DOI and to the Google Scholar link. For the former reference, unfortunately the DOI link does not work well. Instead of pointing to the correct article's webpage, it points to Nat Commun 16:8, that is to the article we first considered in this subsection, which received the false citation. Notice that the number of pages was used instead of the article number, but also the volume changed. We do not know what could be causing the volume number change.

Finally, note that incorrect reference formats appear for works from other publishers as well, for example, the reference written as
\begin{quote}
    You J, Qi PR, Jia ZJ, Wang Y, Wang D, Tian LL, Qi T (2023) Facile preparation of peanut shell derivatives supported MoS2 nanosheets for hydrogen evolution reaction. Catal Commun 179:8
\end{quote}
is published by Elsevier, and in fact has an article number of 106693. Again, the number of pages of the PDF, 8, entered the Reference. For articles that use page numbers and not Article Numbers, the formatting of references appears to be appropriate; it is only Article Numbers which are improperly handled. This also shows that the issue does not only stem from the API of SpringerLink, but also from the automatic formatting of references.

\subsection{Example of O.3}
Consider the article~\cite{wayback_2025_O3example}
\begin{quote}
    Sorour, S.E., Aljaafari, M., Shaker, A.M. et al. LSTM-JSO framework for privacy preserving adaptive intrusion detection in federated IoT networks. Sci Rep 15, 11321 (2025). \href{https://doi.org/10.1038/s41598-025-95966-z}{https://doi.org/10.1038/s41598-025-95966-z}
\end{quote}
which cites the article
\begin{quote}
    30. Chen, X. \& Liu, Y. A comprehensive review of ai-based intrusion detection systems in iot. IEEE Transactions on Emerging Topics in Computational Intelligence 8, 22–39 (2024).
\end{quote}
There are three links provided: CAS, MATH and Google Scholar, which is already suspicious, as CAS is a division of the American Chemical Society, and zbMATH is ``the world's most comprehensive and longest-running abstracting and reviewing service in pure and applied mathematics.''. All three links point to different articles. \href{https://www.nature.com/articles/cas-redirect/1:CAS:528:DC%2BB2cXmtVymsLs%3D}{CAS} points to
\begin{quote}
    Chen, Haiyan et al. Physical activity and exercise in liver cancer. Liver Research, 8(1), 22-33 (2024),
\end{quote}
\href{http://www.emis.de/MATH-item?1313.94082}{MATH} points to
\begin{quote}
    Liu, Jian; Chen, Lusheng. On nonlinearity of the second type of multi-output Boolean functions. Chin. J. Eng. Math. 31 (1), 9-22 (2014).
\end{quote}
and the \href{http://scholar.google.com/scholar_lookup?&title=A%20comprehensive%20review%20of%20ai-based%20intrusion%20detection%20systems%20in%20iot&journal=IEEE%20Transactions%20on%20Emerging%20Topics%20in%20Computational%20Intelligence&volume=8&pages=22-39&publication_year=2024&author=Chen%2CX&author=Liu%2CY}{Google Scholar} to
\begin{quote}
    Sowmya T., Mary Anita E.A. A comprehensive review of ai-based intrusion detection systems in iot. Measurement: Sensors, Volume 28, 100827 (2023).
\end{quote}
Notice that while this last one points to an article with the same title as the one written in the References section, the authors and journal are different.

In the quest to find the true reference (Chen, X. \& Liu, Y., IEEE Transactions on Emerging Topics in Computational Intelligence 8, 22–39 (2024)), we searched based on the volume and page numbers directly on the IEEE Transactions on Emerging Topics in Computational Intelligence website. We find that it does not exist. Articles in Volume 8 have page numbers 16-31 and 32-43, but there is no article that has these numbers.

Finally, note that there are other references for example in this paper which have similar issues, e.g. Reference
\begin{quote}
    28. Smith, J. \& Brown, A. Optimized deep learning intrusion detection for iot security. Nature Communications 15, 1–15 (2024).
\end{quote}
We simply could not find this article anywhere (not even an article with this title), and the links point to different articles, again to CAS, MATH and Google Scholar. Checking even more hyperlinks in the References section we found many to lead to improper webpages.

\newpage
\section{Journals we tested}\label{app:journals}
We explicitly looked at a sample article for each of the following journals, the first article from April 2025, and found no Article Number in the API responses, except for the JATS format response, as detailed in I.2. The journals are Nature Communications, Scientific Reports, BMC Public Health, BMC Medicine, Genome Biology, Molecular Cancer, BMC Genomics, Communications Biology, Communications Chemistry, Communications Physics, Communications Engineering, Communications Medicine, Communications Earth \& Environment, npj Quantum Information, npj Computational Materials, npj Clean Water, npj Digital Medicine, npj Microgravity, npj Materials Degradation, npj Biofilms and Microbiomes, BMC Biology, BMC Neuroscience, Cell Death \& Disease, Light: Science \& Applications, Discover Materials, Discover Biotechnology, Discover Applied Sciences. We have not explicitly examined histograms or rankings for all these journals, but presumably, as the technical issue seems to be present, they also have distorted citation metrics.

\newpage
\section{Citation count histograms}\label{app:histograms}
\begin{figure*}[h!]
    \centering
    \includegraphics[width=0.32\linewidth]{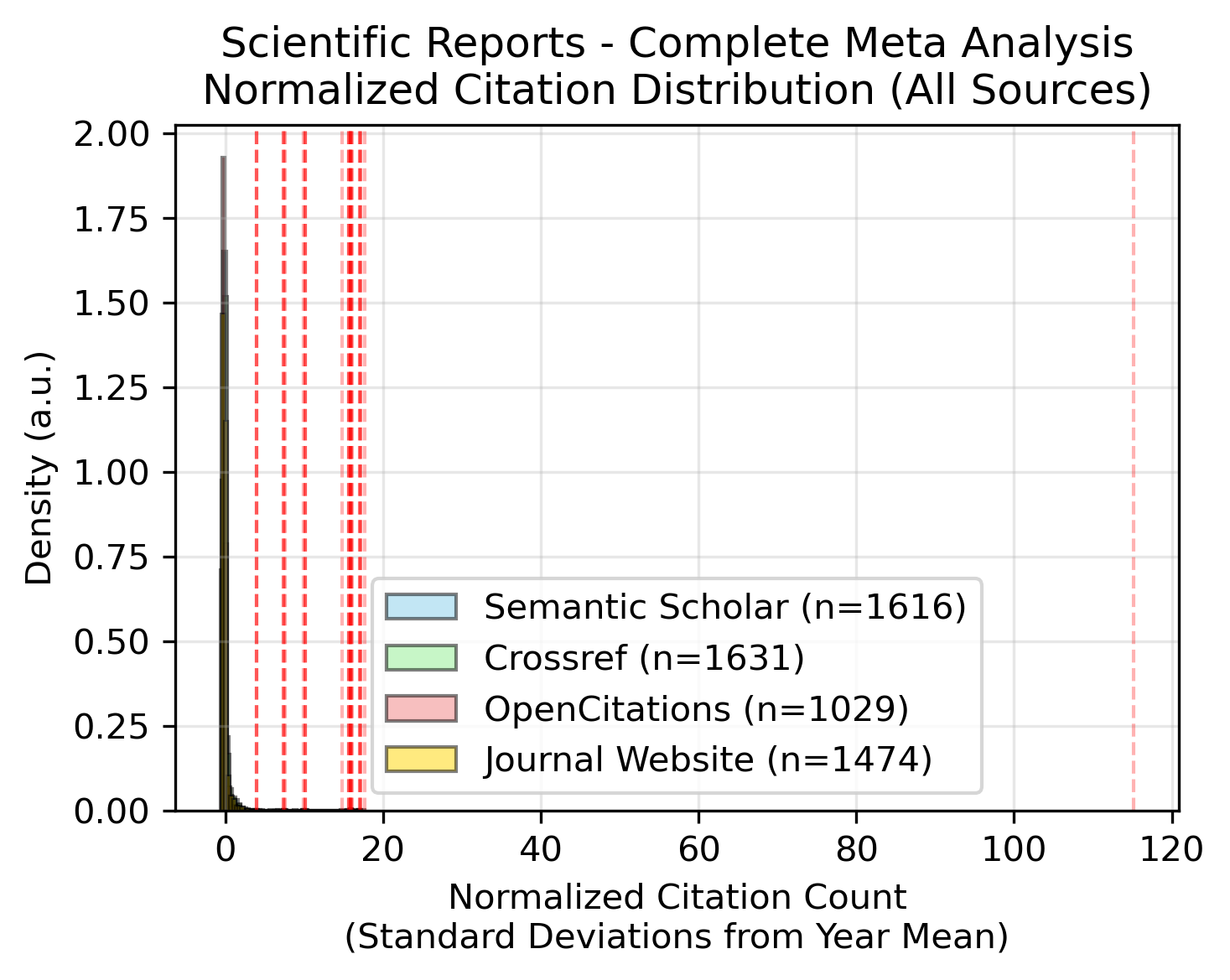}
    \includegraphics[width=0.32\linewidth]{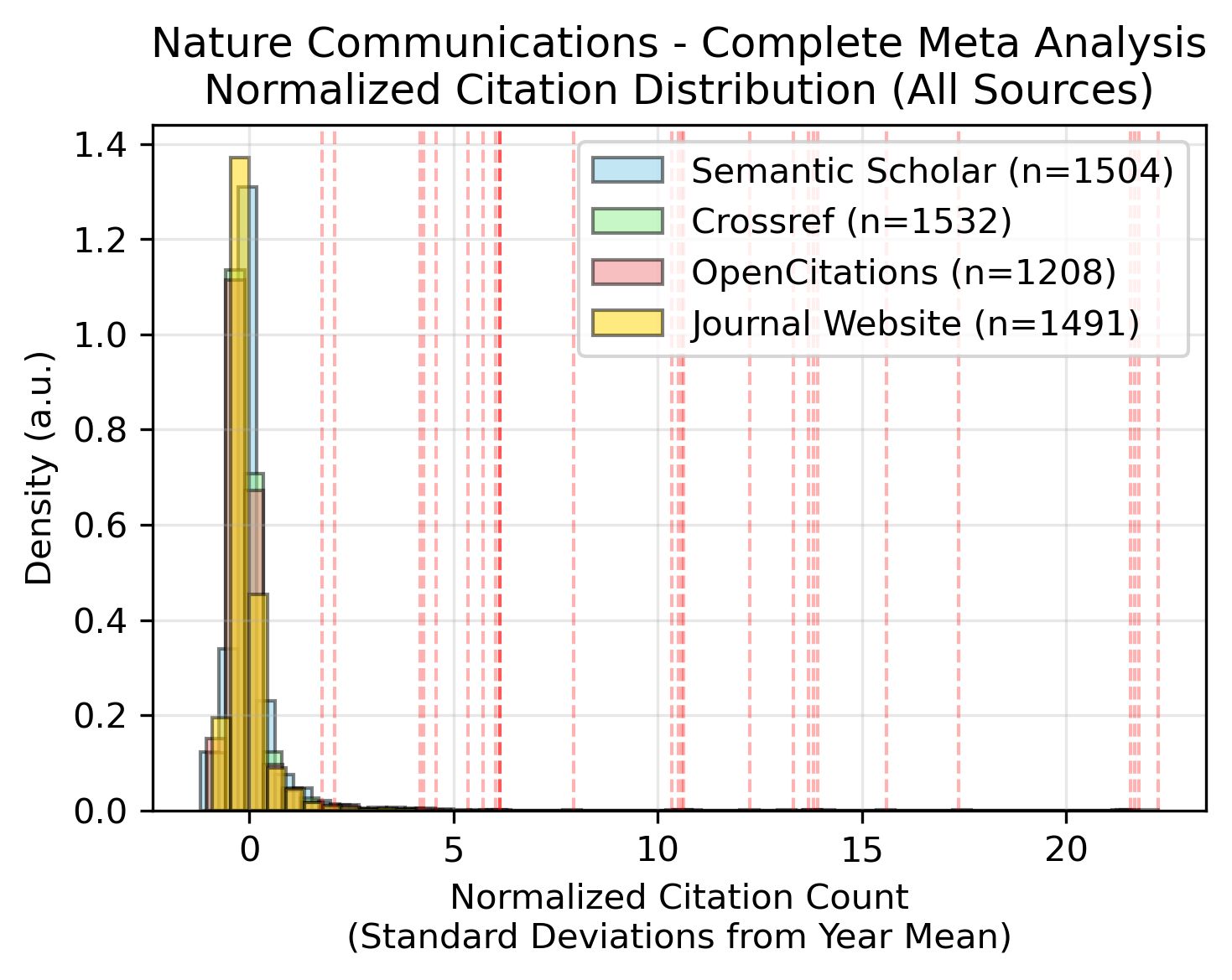}
    \includegraphics[width=0.32\linewidth]{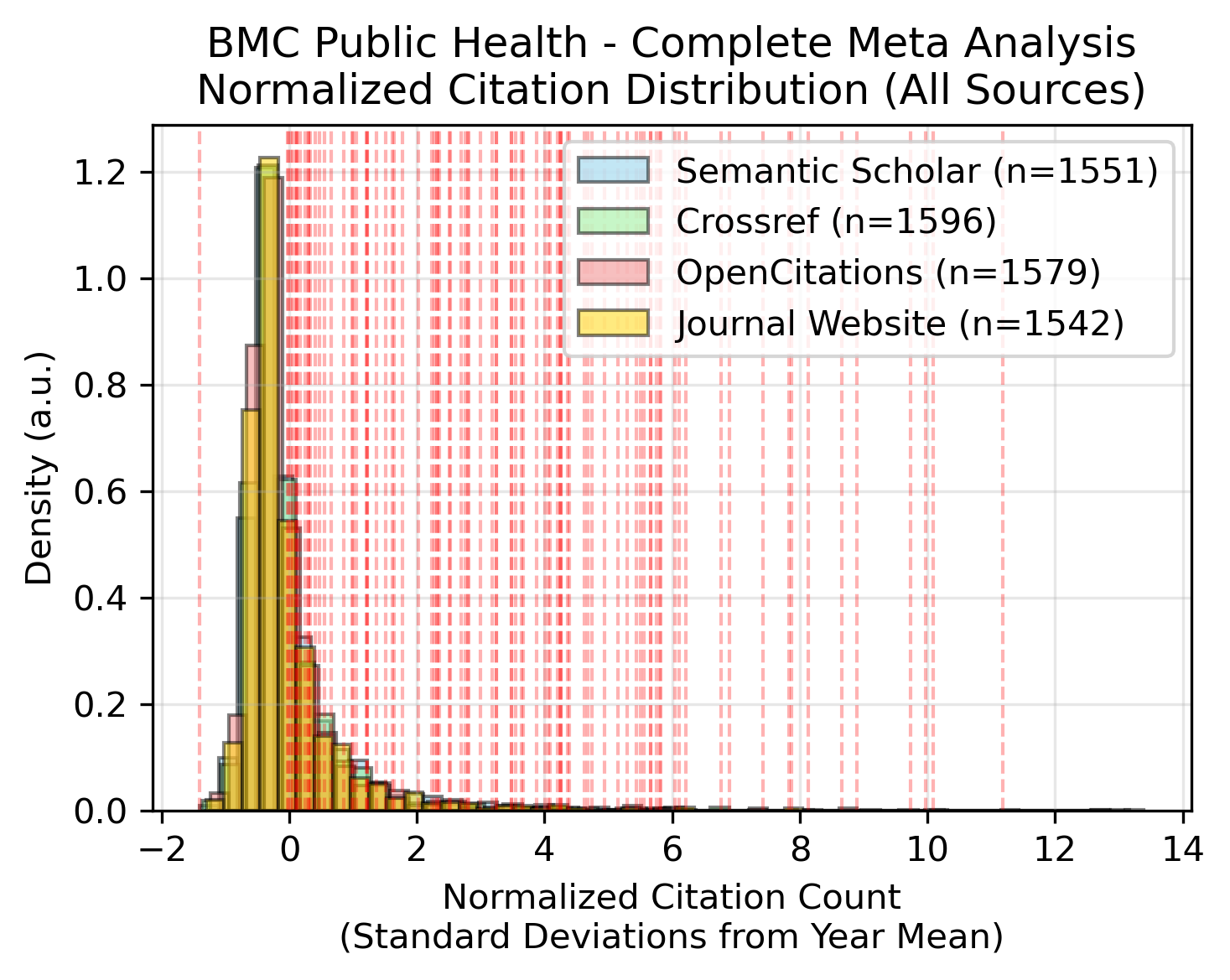}
    \caption{Normalized citation count histogram for a) Scientific Reports b) Nature Communications and c) BMC Public Health aggregated across all years and sources. For BMC Public Health the only Article 1 below the mean had 0 citations due to the technical issue we had with OpenCitations for the year 2025.}
    \label{fig:figAppB_1}
\end{figure*}

\begin{figure*}[t!]
    \centering
    \includegraphics[width=\linewidth]{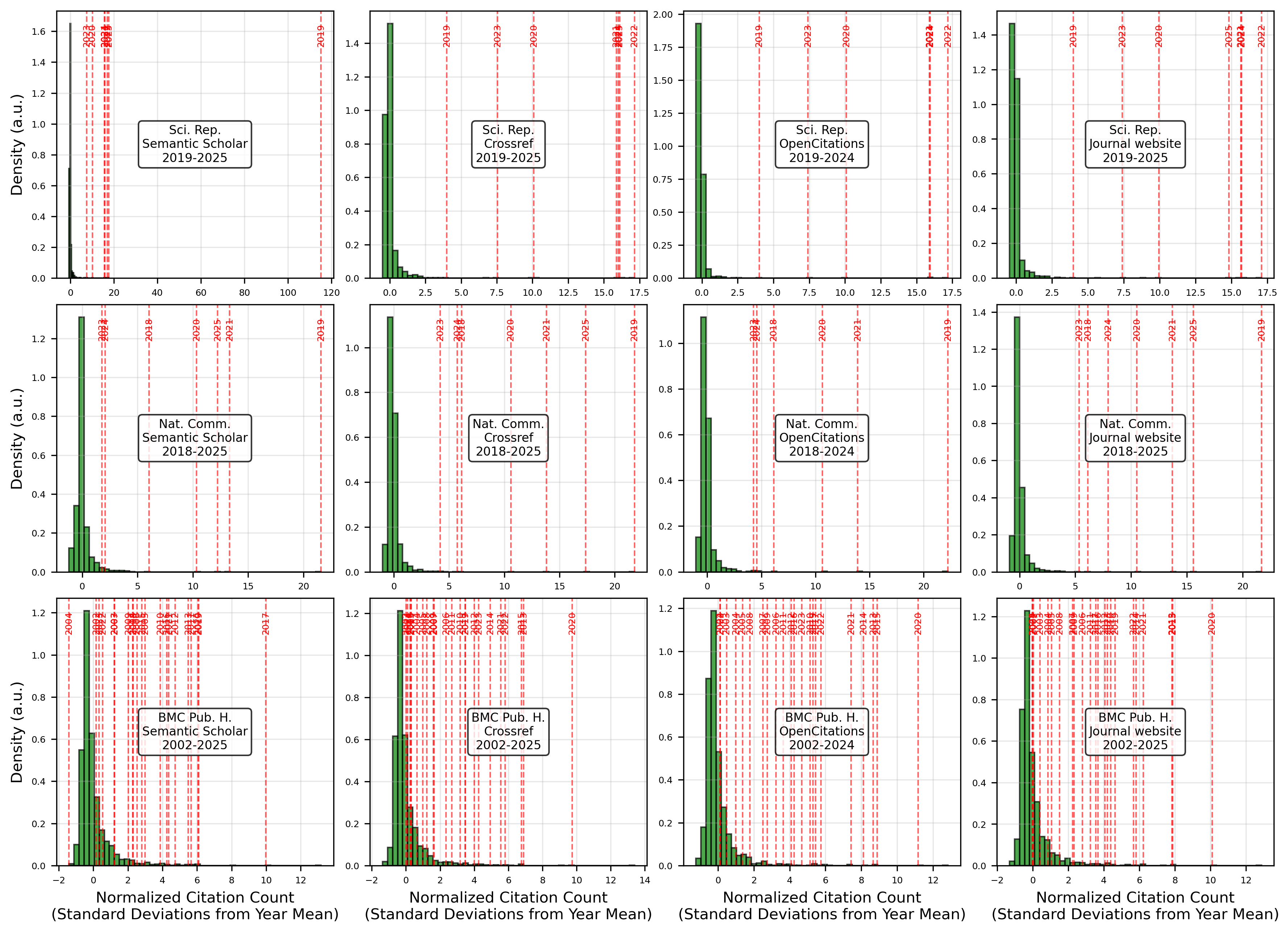}
    \caption{Normalized citation count histogram for Scientific Reports, Nature Communications and BMC Public Health for each citation count source, aggregated across all years.}
    \label{fig:figAppB_2}
\end{figure*}

\begin{figure*}[b]
    \centering
    \includegraphics[width=0.48\linewidth]{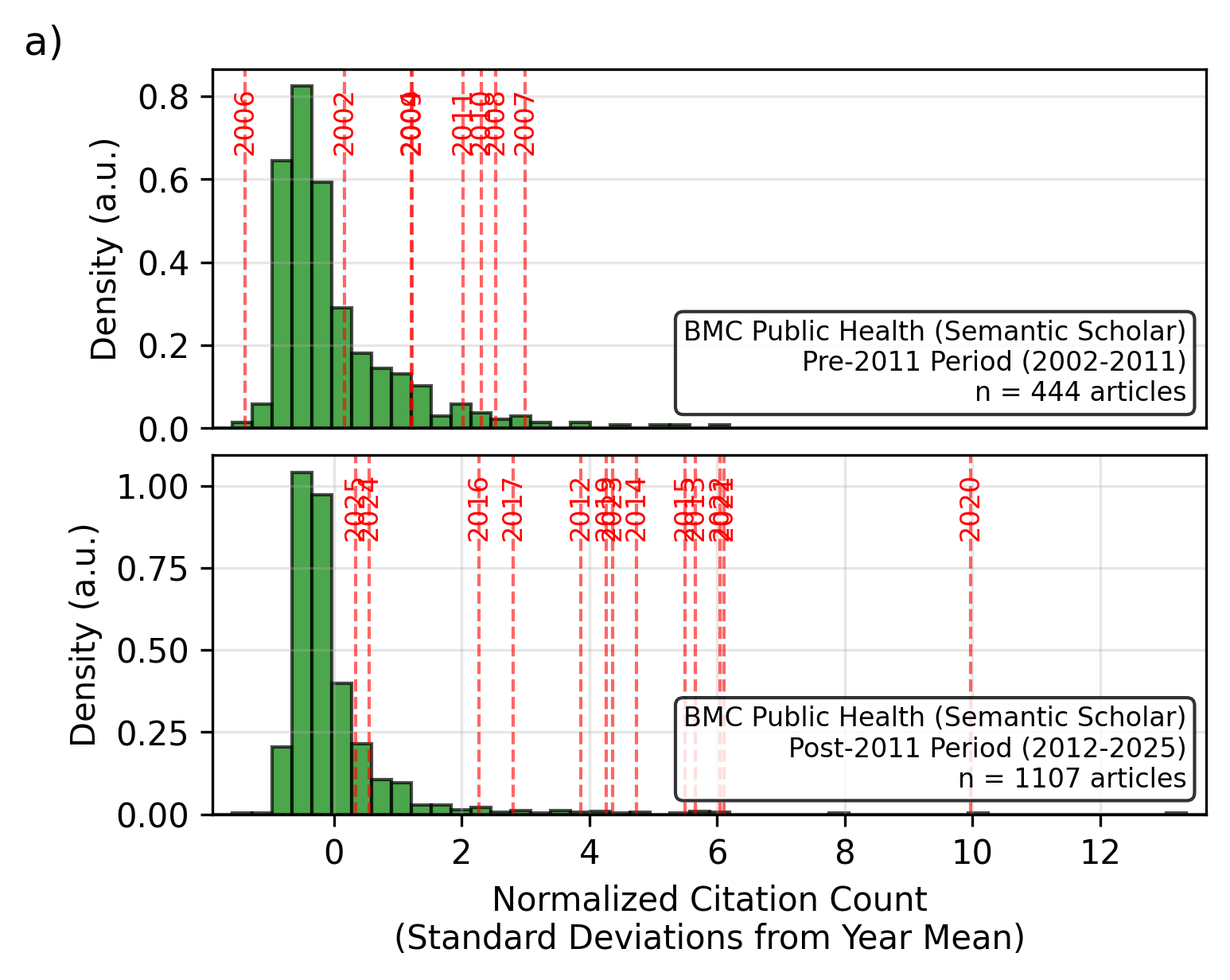}
    \includegraphics[width=0.48\linewidth]{bmc_public_health_crossref_split_histograms_no_title.png}
    \includegraphics[width=0.48\linewidth]{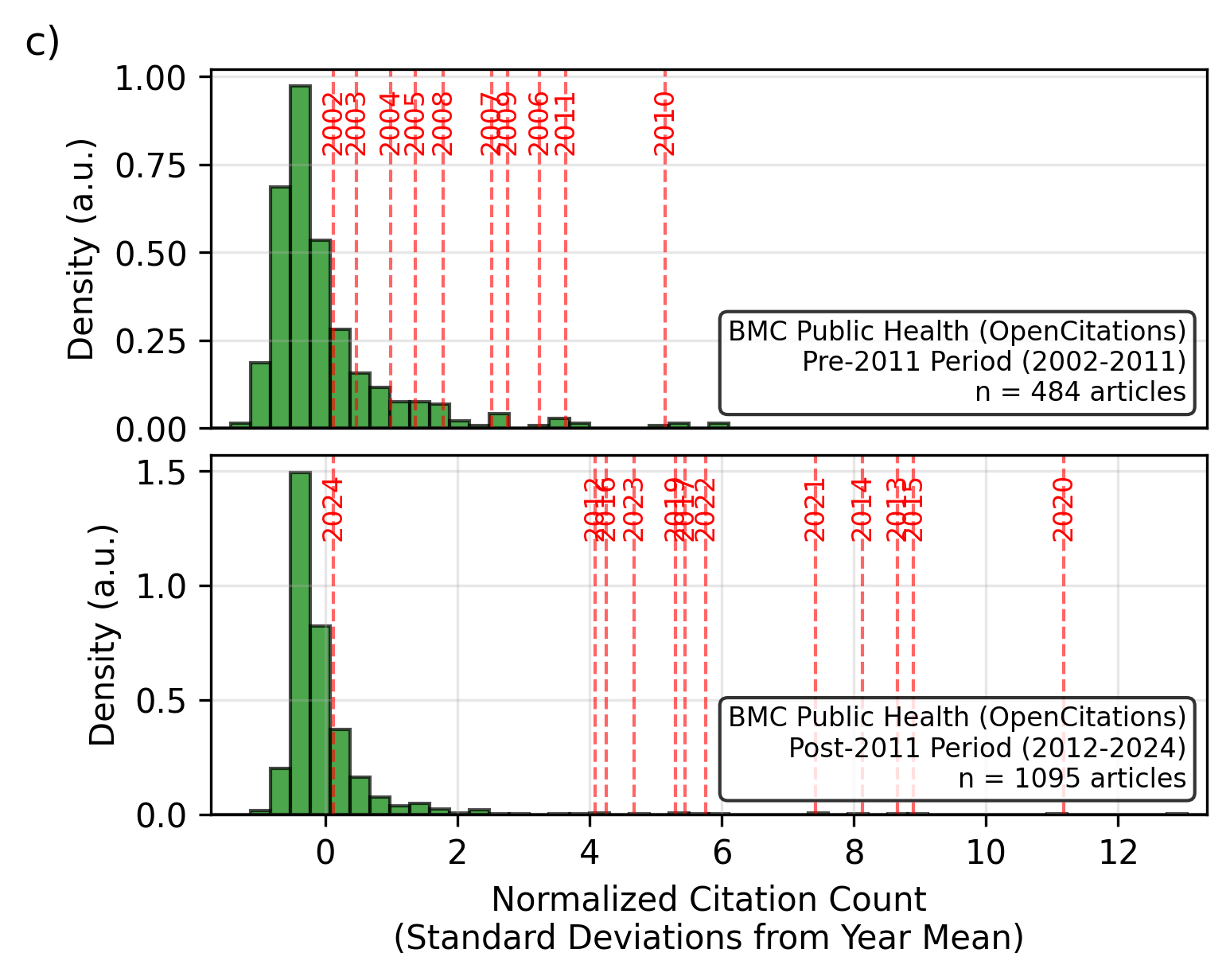}
    \includegraphics[width=0.48\linewidth]{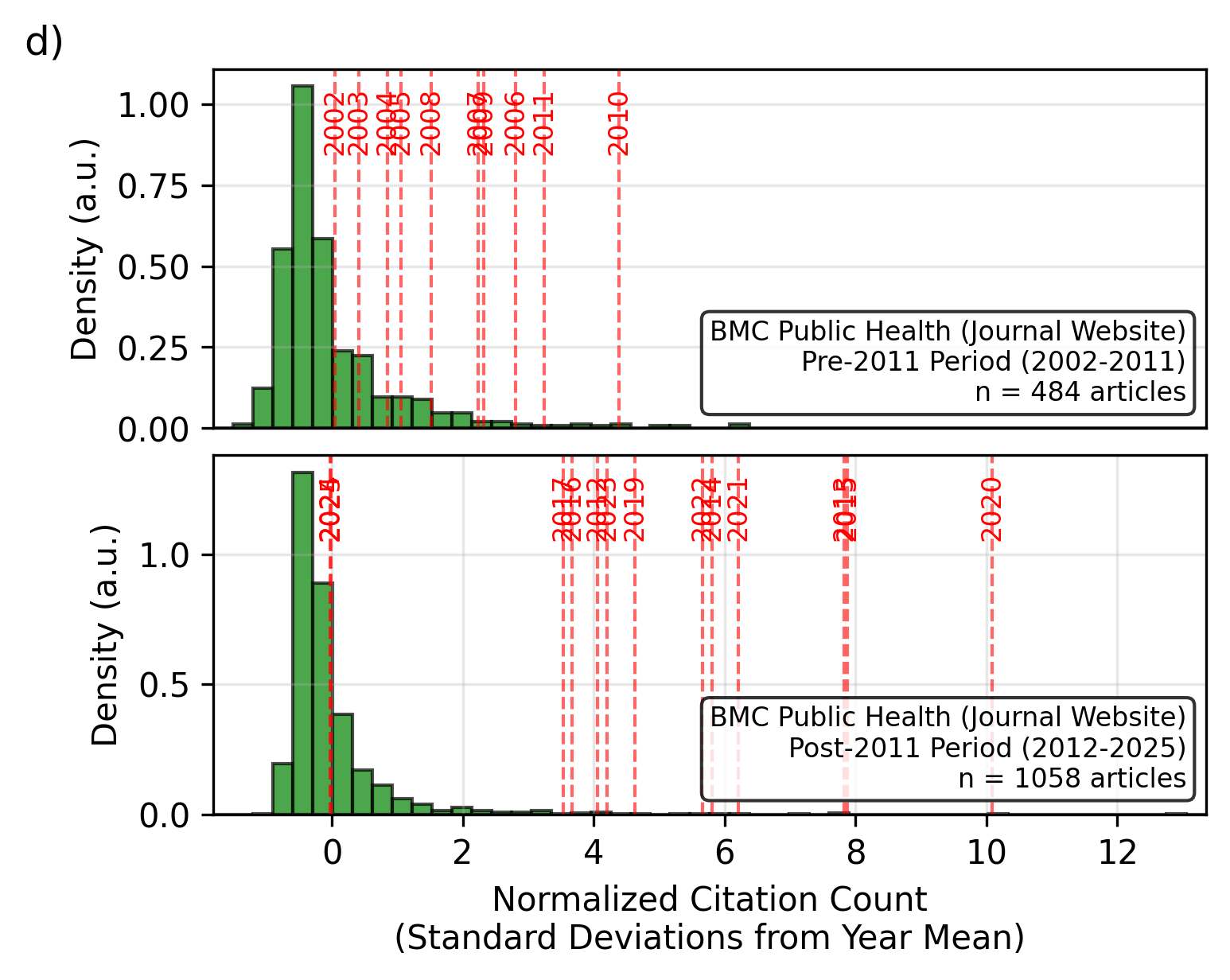}
    \caption{Normalized citation count histograms for BMC Public Health aggregated across all years for citation count sources a) Semantic Scholar b) Crossref c) OpenCitations d) journal website, split for years before 2011 and after 2011. These temporally split histograms show that the issue may have persisted since 2011, the year the API was introduced.}
    \label{fig:figAppB_3}
\end{figure*}

We first give aggregate histograms, where we aggregate data over all years and citation count sources in Fig.~\ref{fig:figAppB_1}. Then in Fig.~\ref{fig:figAppB_2} we show the same data, broken down for each citation count source (aggregated for all years). Next, we provide temporally split histograms for each citation count source in Fig.~\ref{fig:figAppB_3}. On the pages thereafter we give year-by-year citation count histograms for the following journals: Scientific Reports, Nature Communications, BMC Public Health.

For the analysis, we use four sources of data: Crossref, Semantic Scholar, OpenCitations and the citation count on the journals' own websites. Note that for Nature Communications, 2022 was left out as Article 1 was a correction. Several technical issues we did not resolve: for BMC Public Health our script had difficulties finding Article 1 for the first year, 2001. For the OpenCitations client, for the moment we could not get citation count through the API for 2025. We did not take the time to fix this, and instead worked with the other 24 years (2002-2025) for BMC and with the other data sources for 2025. The year 2025 in all cases spans up to approximately 2nd or 3rd of October 2025, as data was collected at this time.

\begin{figure}[t!]
    \centering
    \includegraphics[width=\linewidth]{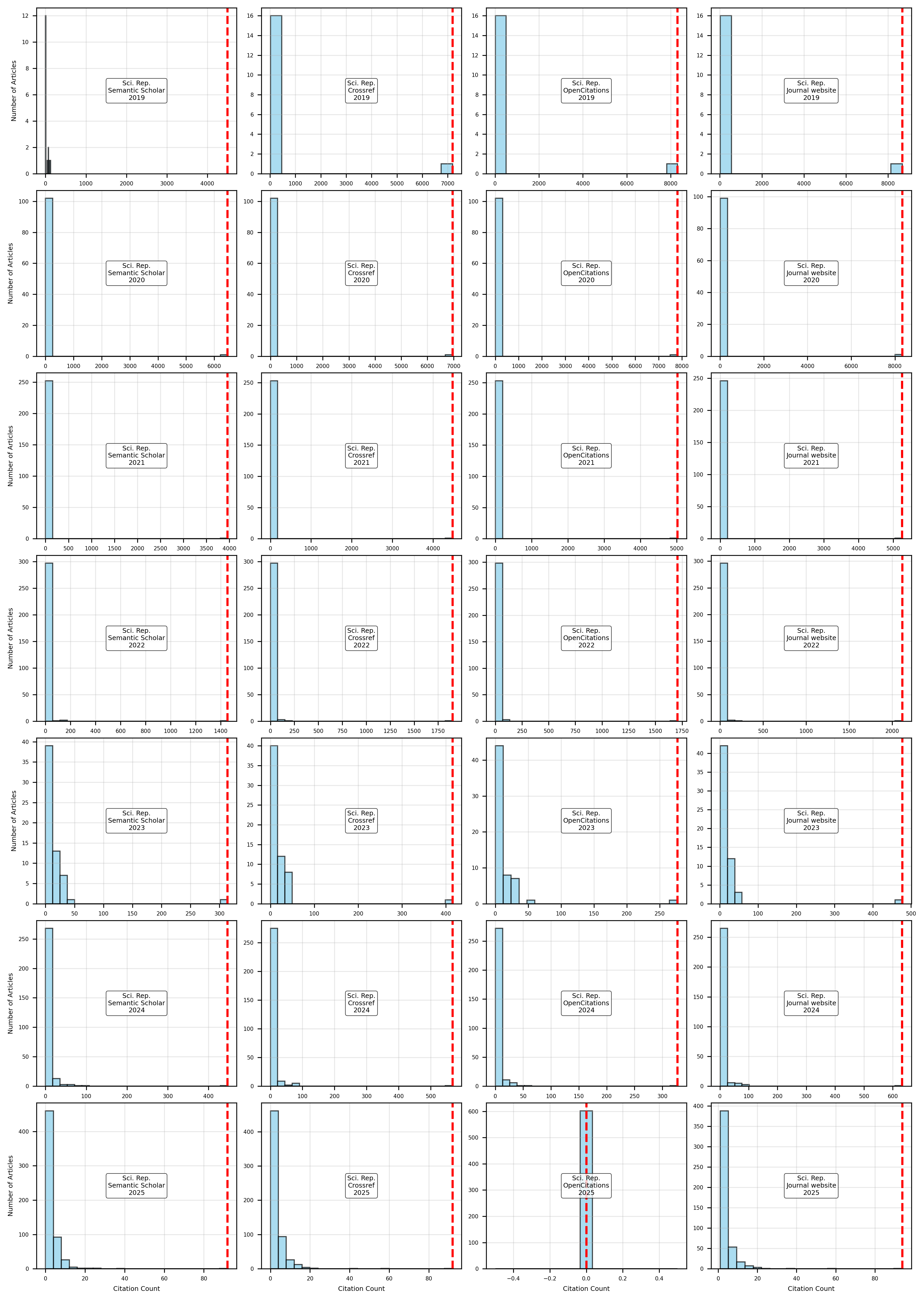}
\end{figure}

\begin{figure}[t!]
    \centering
    \includegraphics[width=\linewidth]{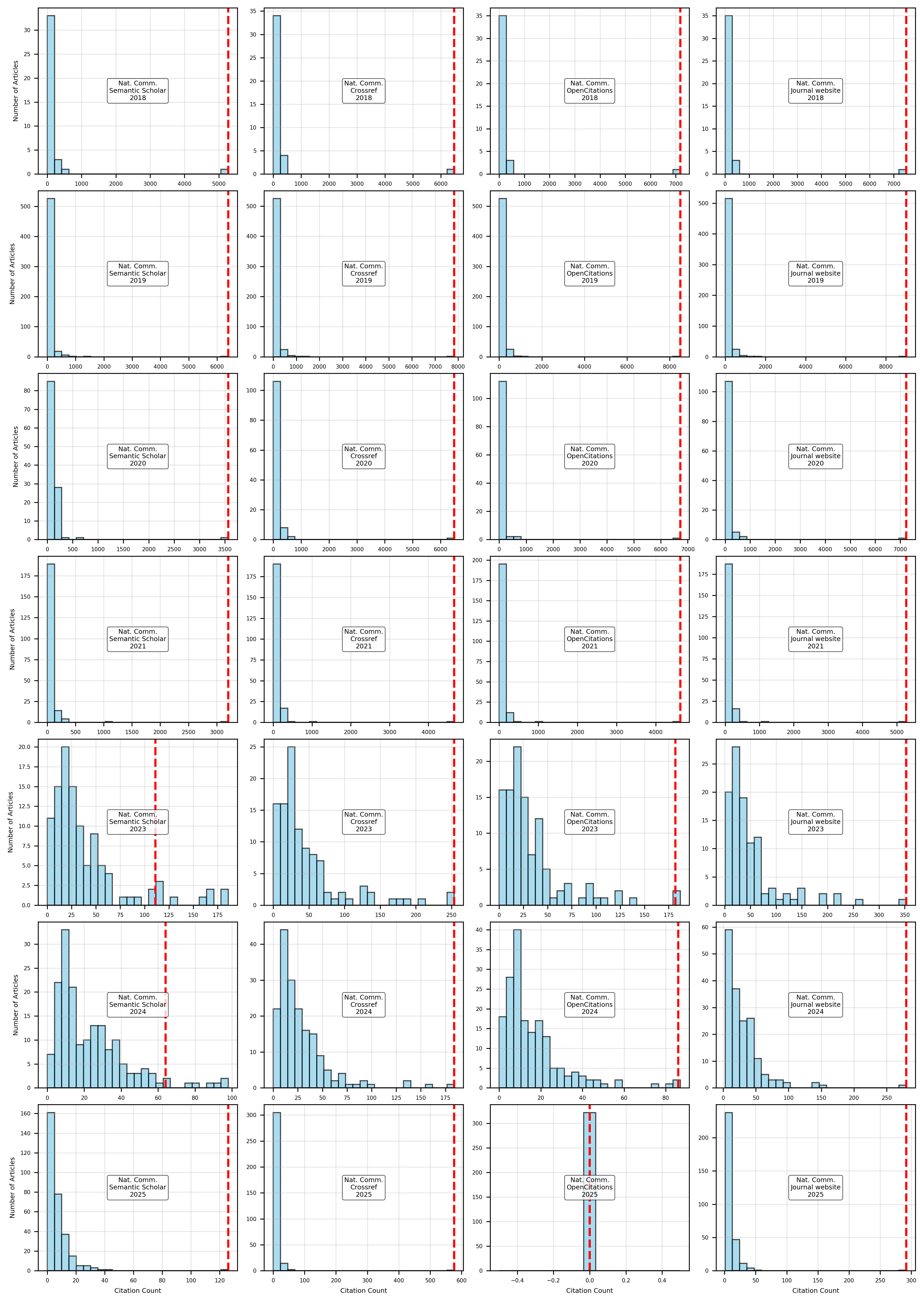}
\end{figure}

\begin{figure}[t!]
    \centering
    \includegraphics[width=\linewidth]{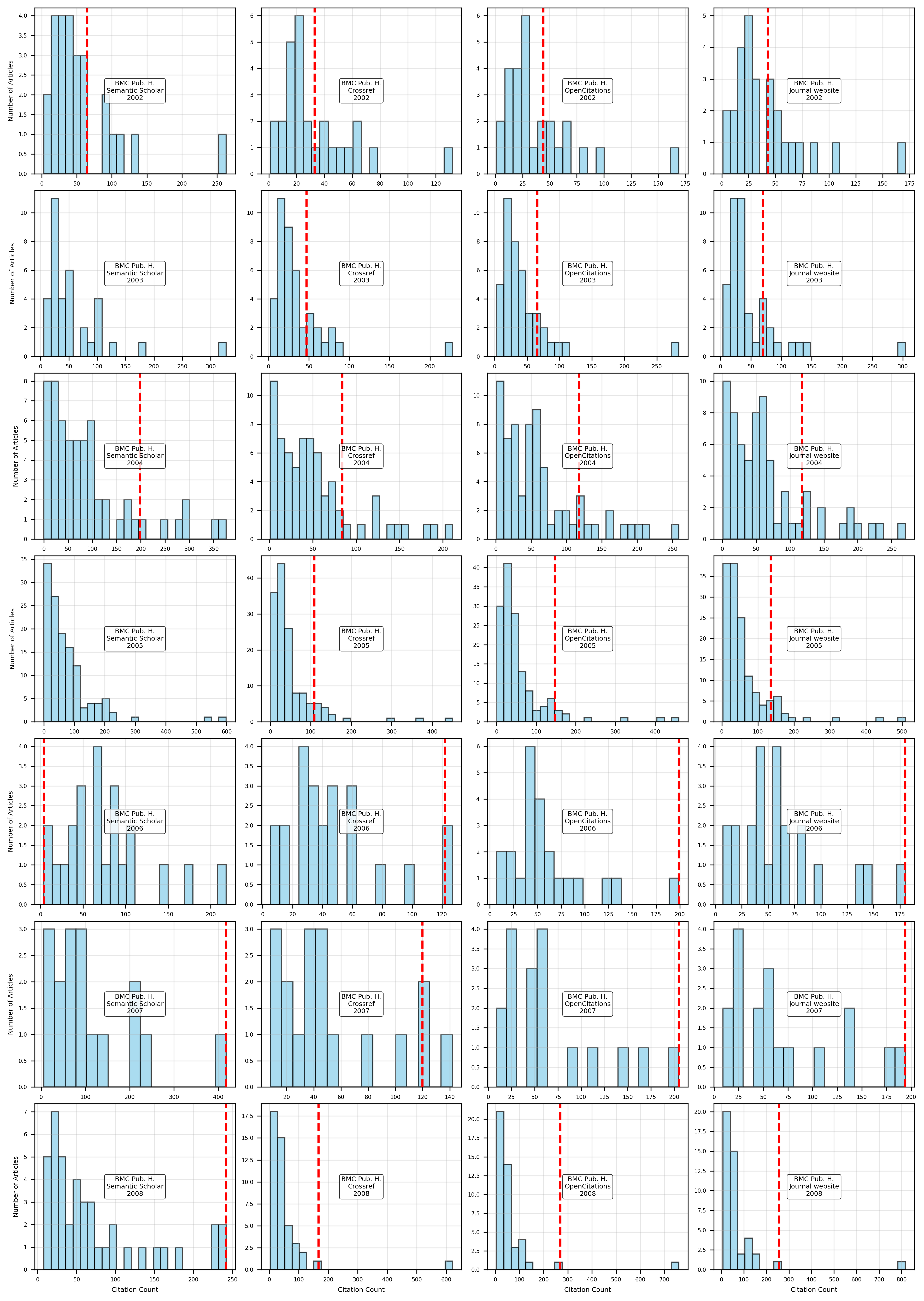}
\end{figure}

\begin{figure}[t!]
    \centering
    \includegraphics[width=\linewidth]{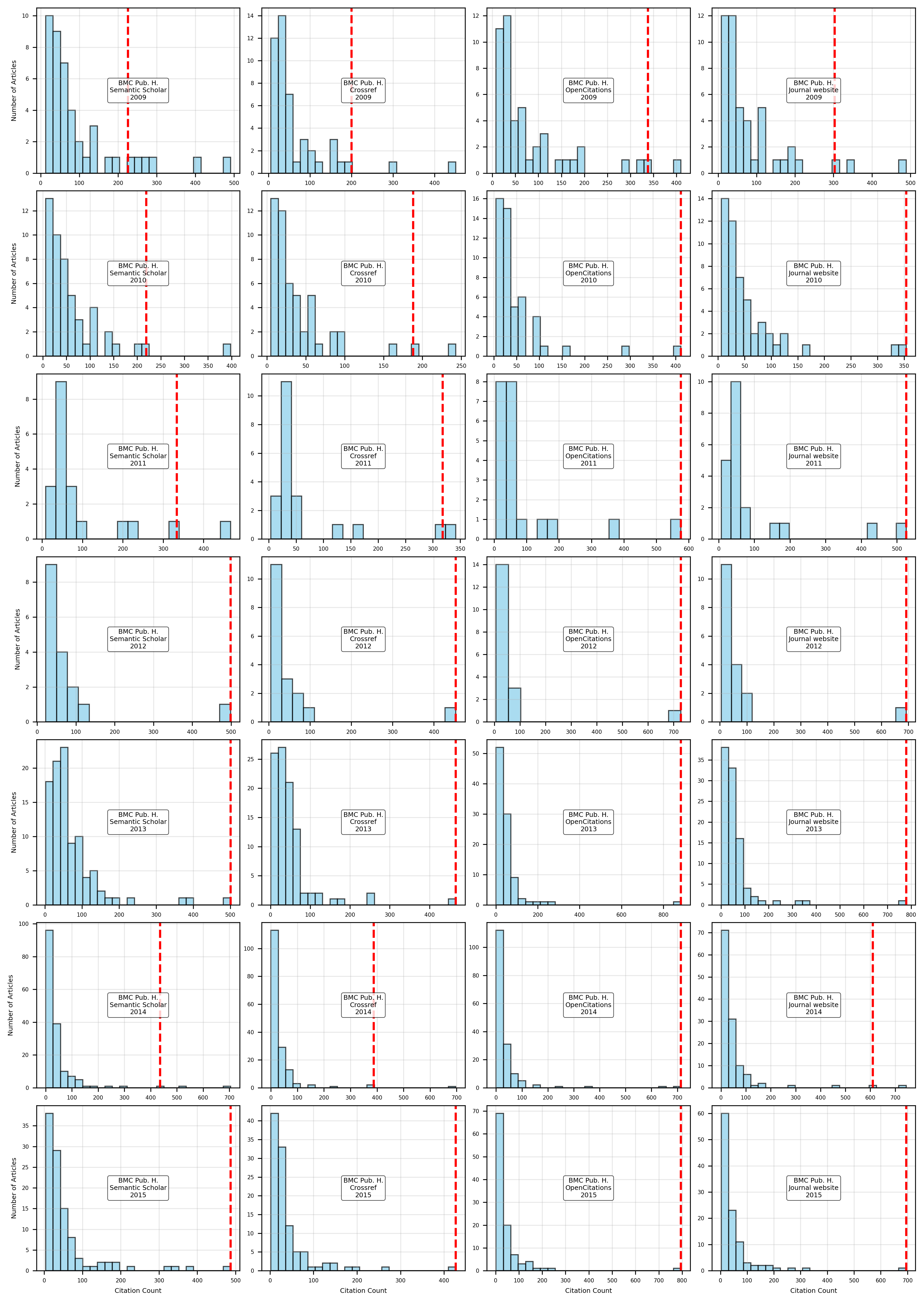}
\end{figure}

\begin{figure}[t!]
    \centering
    \includegraphics[width=\linewidth]{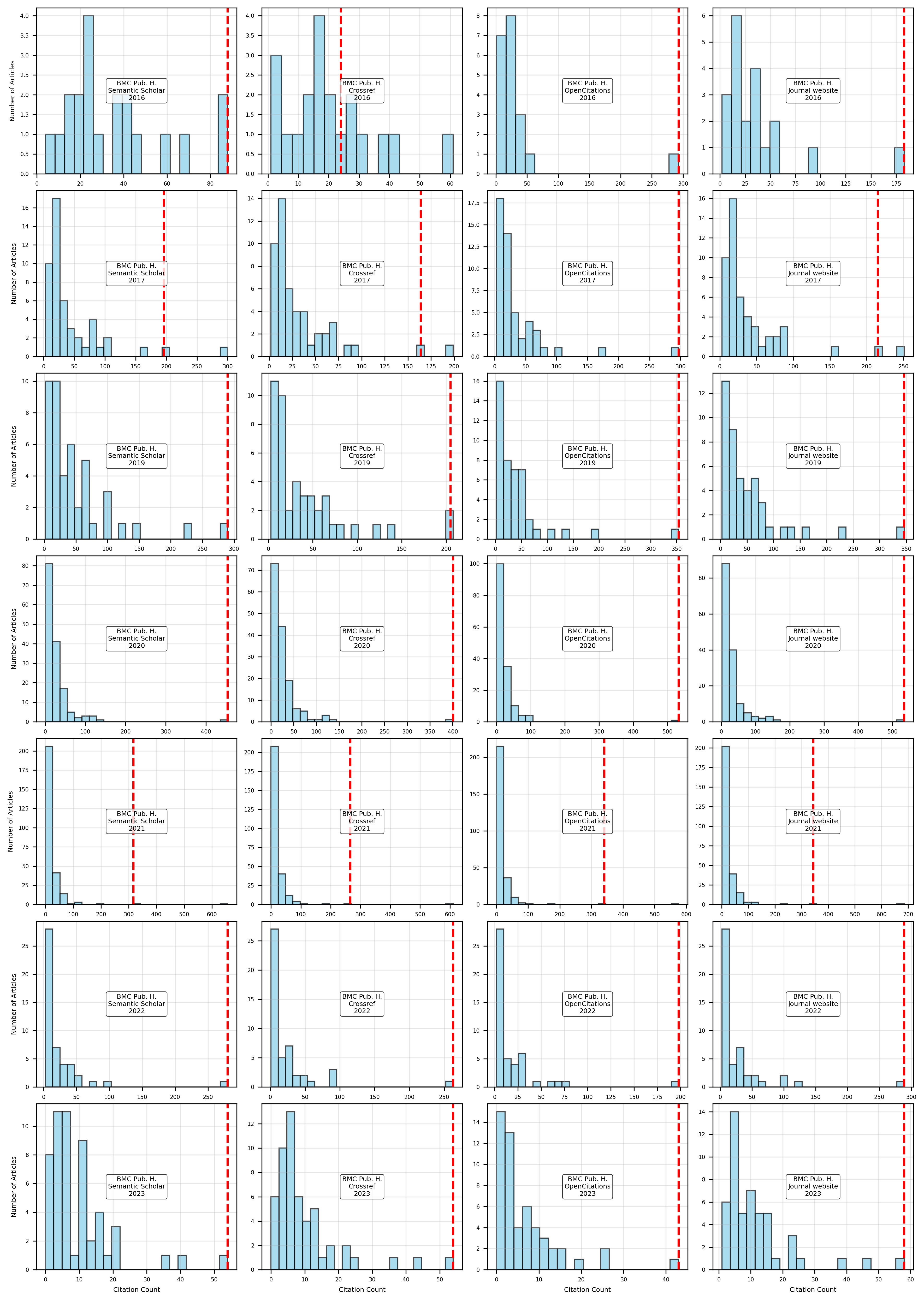}
\end{figure}

\begin{figure}[t!]
    \centering
    \includegraphics[width=\linewidth]{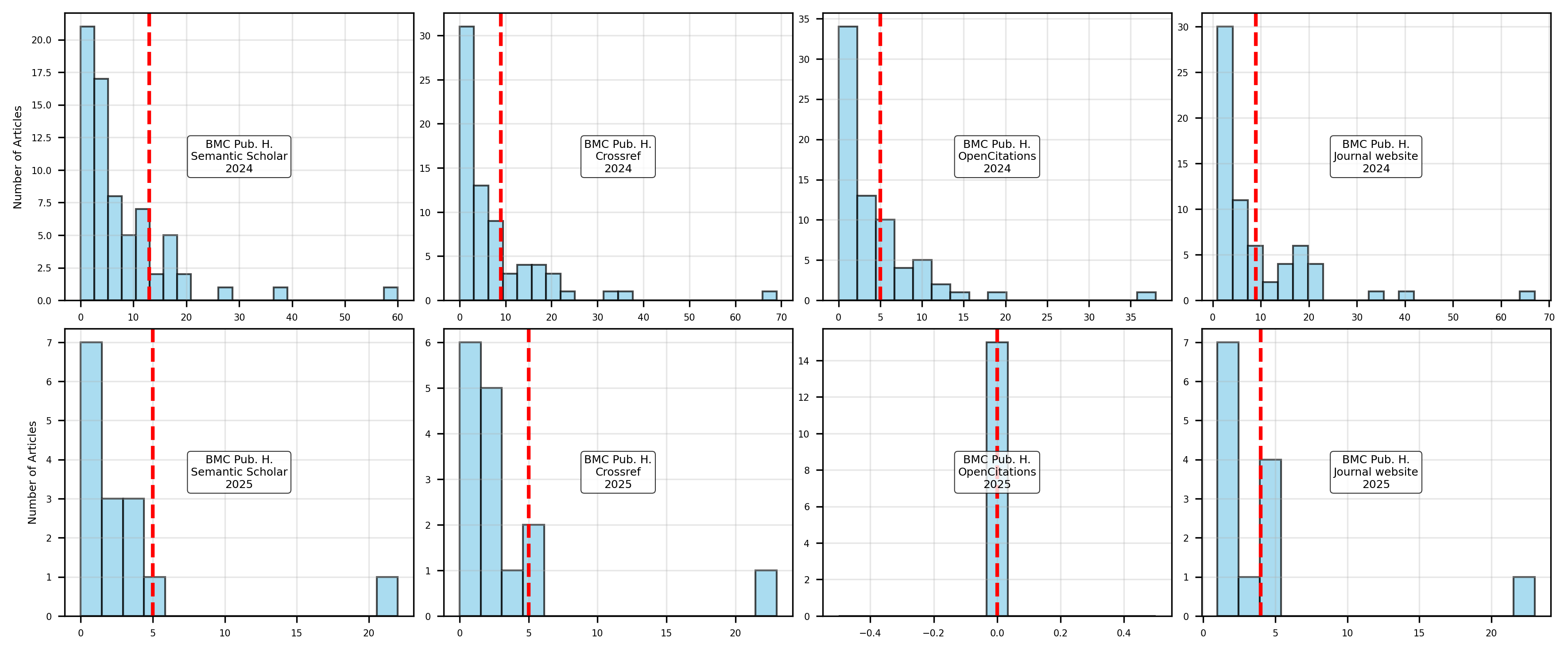}
\end{figure}

\end{document}